\documentclass{aa}
\usepackage[varg]{txfonts}
\usepackage{amsmath}
\usepackage{amssymb}
\usepackage{booktabs}
\usepackage{graphicx}
\usepackage[normal]{caption}
\usepackage[labelfont=bf]{caption}
\usepackage{subcaption}
\usepackage{tablefootnote}
\usepackage{threeparttable}
\usepackage{natbib}
\usepackage[usenames,dvipsnames]{xcolor}
\usepackage[bookmarks, colorlinks, breaklinks]{hyperref}
\usepackage{enumitem} 
\usepackage{makecell}
\usepackage{colortbl}
\usepackage{tabularx}
\usepackage{upgreek}
\usepackage{stfloats}
\hypersetup{linkcolor=blue,citecolor=blue,filecolor=black,urlcolor=blue}
\bibpunct{(}{)}{;}{a}{}{,} 

\begin{document}

\title{Testing baryon-induced core formation in $\Lambda$CDM: A comparison of the DC14 and coreNFW dark matter halo models on galaxy rotation curves}
\titlerunning{Testing baryon-induced core formation in $\Lambda$CDM}
\author{F.~Allaert\inst{1}
\and G.~Gentile\inst{2}
\and M.~Baes\inst{1}}
\institute{Sterrenkundig Observatorium, Universiteit Gent, Krijgslaan 281, B-9000 Gent, Belgium\\ \email{Flor.Allaert@UGent.be}
\and Department of Physics and Astrophysics, Vrije Universiteit Brussel, Pleinlaan 2, 1050 Brussels, Belgium
}
\date{}
\abstract{
Recent cosmological hydrodynamical simulations suggest that baryonic processes, and in particular supernova feedback following bursts of star formation, can alter the structure of dark matter haloes and transform primordial cusps into shallower cores. To assess whether this mechanism offers a solution to the long-standing cusp-core controversy, simulated haloes must be compared to real dark matter haloes inferred from galaxy rotation curves. For this purpose, two new dark matter density profiles were recently derived from simulations of galaxies in complementary mass ranges: the DC14 halo ($10^{10} < M_{\text{halo}}/M_{\odot} < 8 \times 10^{11}$) and the \textsc{core}NFW halo ($10^{7} < M_{\text{halo}}/M_{\odot} < 10^{9}$). Both models have individually been found to give good fits to observed rotation curves. For the DC14 model, however, the agreement of the predicted halo properties with cosmological scaling relations was confirmed by one study, but strongly refuted by another.\\
A next important question is whether, despite their different approaches, the two models converge to the same solution in the mass range where both should be appropriate. To investigate this, we tested the DC14 and \textsc{core}NFW halo models on the rotation curves of a selection of galaxies with halo masses in the range $4 \times 10^{9}$ $M_{\odot}$ - $7 \times 10^{10}$ $M_{\odot}$ and compared their predictions. We further applied the DC14 model to a set of rotation curves at higher halo masses, up to $9 \times 10^{11}$ $M_{\odot}$, to verify the agreement with the cosmological scaling relations.\\
Both models are generally able to reproduce the observed rotation curves, in line with earlier results, and the predicted dark matter haloes are consistent with the cosmological $c-M_{\text{halo}}$ and $M_{*}-M_{\text{halo}}$ relations. We find that the DC14 and \textsc{core}NFW models are also in fairly good agreement with each other, even though DC14 tends to predict slightly less extended cores and somewhat more concentrated haloes than \textsc{core}NFW. While the quality of the fits is generally similar for both halo models, DC14 does perform significantly better than \textsc{core}NFW for three galaxies. In each of these cases, the problem for \textsc{core}NFW is related to connection of the core size to the stellar half-mass radius, although we argue that it is justifiable to relax this connection for NGC\,3741. A larger core radius brings the \textsc{core}NFW model for this galaxy in good agreement with the data and the DC14 model.
}
\maketitle

\section{Introduction}

For several decades the dark matter problem has been one of the main topics in astronomical research. The  idea of missing or invisible mass was already proposed in the early 1930s by Jan Oort \citep{oort32} and Fritz Zwicky \citep{zwicky33} based on their observations of the motions of stars in the Milky Way disk and galaxies in the Coma cluster. Despite this early notion, the first sound evidence of the presence of dark matter only came in the 1970s, from the analysis of galaxy rotation curves by  \cite{freeman70}, \cite{roberts75}, and \cite{rubin78}. These authors found that the rotation curves of massive galaxies remain flat even at large galactocentric distances and well beyond the stellar disks. This could not be explained by the Newtonian gravity of the visible matter alone, but instead implied an additional extended halo of invisible matter. Furthermore it was found that the rotation curves of low mass and low surface brightness (LSB) galaxies show a slow, almost linear rise in the centre after subtraction of the baryonic contributions. To match this observed behaviour, empirical models of the dark matter distribution in galaxies typically have a central constant-density core. These models, such as the pseudo-isothermal sphere, can explain a wide variety of observed rotation curves \citep[e.g.][]{vanalbada85, broeils92, deblok01}, although they have no physical basis.
\par
On the other hand, dark matter only simulations of the structure formation in the Universe consistently find dark matter haloes with a central cusp \citep[e.g.][]{navarro96, moore99, klypin01, diemand05, stadel09}. The dark matter density profiles derived from these simulations, however, give poor fits to the rotation curves of dwarf galaxies \citep[e.g.][]{deblok01, deblok02, weldrake03, gentile05, gentile07}. This cusp-core controversy has been one of the major problems of $\Lambda$CDM for the past two decades. In the early years rotation curves were often derived in an overly simplistic way from poorly sampled velocity fields or even one-dimensional long-slit observations. Because both physical effects, such as non-circular motions and pressure support, and observational biases, such as beam smearing, often affect the observed kinematics in the central parts of galaxies, there has long been a discussion regarding whether the observed rotation curves are actually reliable and truly trace the gravitational potential of a galaxy \citep[e.g.][]{swaters03,rhee04,spekkens05,valenzuela07}. Since the 1990s, however, both the quality of the observations and the analysis techniques have vastly improved and although the discussion still persists today \citep{pineda16}, there is growing consensus that modern rotation curves accurately trace the total gravitational potential in a galaxy, at least for properly selected systems. Despite these improvements and the substantially higher resolution of present-day dark matter only simulations, the discrepancy still persists.
\par
An alternative solution to the cusp-core problem is that baryonic processes associated with galaxy formation and evolution also affect the dark matter halo. Various processes have been proposed in this context with different results. On the one hand, condensation of cooling gas towards the centre of a galaxy causes a further contraction of the dark matter halo and a stronger cusp \citep[e.g.][]{jesseit02, gnedin04}. On the other hand, infalling gas clumps can transfer angular momentum to the dark matter via dynamical friction, ultimately resulting in a shallower central profile \citep{elzant01,tonini06,romano08}, but the efficiency of this mechanism is still under debate \citep[e.g.][]{deblok10}. Finally, feedback from supernovae (and AGN activity in high mass galaxies) can induce massive gas outflows that also cause the dark matter halo to expand. \cite{navarro96b} already investigated this scenario using a highly simplified outflow model and concluded that supernova feedback could have flattened the dark matter cusps in dwarf irregular galaxies, but is unlikely to be effective in more massive systems. In a more detailed study \cite{read05} found that repeated bursts of star formation alternated by epochs of gas (re-)accretion can indeed gradually transform dark matter cusps into cores in simulated dwarf galaxies. This result was later confirmed and extended to somewhat higher mass galaxies by numerous studies \citep[e.g.][]{governato10,maccio12,teyssier13,dicintio14a,onorbe15}, although the details of the conclusions sometimes differ. For example, while \cite{governato12} found that supernova feedback can only expand dark matter haloes in galaxies with $M_{*} \gtrsim 10^{7}$ $M_{\odot}$, \cite{read16a} concluded that cores also form in lower mass systems if star formation proceeds for long enough.
\par
Hydrodynamical simulations therefore seem to suggest that stellar feedback effects can solve the cusp-core controversy. To really confirm this claim, however, the simulated haloes must be compared to real observed rotation curves. For this purpose, two new analytic dark matter density profiles were recently proposed. The DC14 profile was derived by \cite{dicintio14b} from their simulated galaxies in the mass range $9.94 \times 10^{9}$ < $M_{\text{halo}}/M_{\odot}$ < $7.8 \times 10^{11}$ and was recently tested on samples of observed rotation curves by \cite{katz16} and \cite{pace16}. Both studies concluded that the DC14 profile can indeed reproduce the observed rotation curves. \cite{katz16} also found that the derived dark matter halo parameters are in excellent agreement with the cosmological stellar mass-halo mass and halo mass-concentration relations. \cite{pace16}, on the other hand, concluded the opposite. He found halo masses significantly below the cosmological prediction for galaxies with stellar masses $M_{*} \lesssim 10^{9}$ $M_{\odot}$, and a huge scatter of almost two orders of magnitude in the derived halo concentrations. The \textsc{core}NFW model was derived by \cite{read16a} from simulations of tiny dwarf galaxies in the mass range $M_{\text{halo}} \sim 10^{7} - 10^{9}$ $M_{\odot}$. \cite{read16c} also recently tested this profile on a set of dwarf galaxies extracted from the Little THINGS sample \citep{hunter12,iorio16} with halo masses ranging from a few $10^{8}$ $M_{\odot}$ to about $2 \times 10^{10}$ $M_{\odot}$. These authors also found good fits to the observed rotation curves and good agreement with the stellar mass-halo mass relation.
\par
Although the physical mechanism that drives core formation is essentially the same for both halo models, they follow a somewhat different approach. In the DC14 model the stellar mass is used as a measure for the amount of supernova feedback energy that has become available and the shape of the dark matter halo is fully determined by $M_{*}/M_{\text{halo}}$. In the \textsc{core}NFW model, on the other hand, the core strength is regulated by the total time that the galaxy has been forming stars, while the radial extent of the core is linked to the radial distribution of the stars. 
\par
The \textsc{core}NFW and DC14 models also probe different mass ranges and are in this sense complementary. While it is probably not meaningful to apply the DC14 model to the rotation curves of tiny dwarf galaxies or to extrapolate \textsc{core}NFW to Milky Way-size systems, the two models should both be appropriate for halo masses of the order $5 \times 10^{9}$ $M_{\odot} \lesssim M_{\text{halo}} \lesssim 5 \times 10^{10}$ $M_{\odot}$. Therefore, if DC14 and \textsc{core}NFW both correctly describe dark matter core formation, their predictions should agree in this overlapping mass range.
\par
In this work we apply the \textsc{core}NFW and DC14 models to a selection of 13 galaxy rotation curves with halo masses of $4 \times 10^{9}$ $M_{\odot}$ to $7 \times 10^{10}$ $M_{\odot}$, and compare their predictions. The DC14 halo is further applied to an additional 7 rotation curves with halo masses up to $9 \times 10^{11}$ $M_{\odot}$ to investigate the agreement with the cosmological scaling relations. This paper is organized as follows: in Section \ref{sec:sample} we describe the selection of our sample of rotation curves. The principle of mass modelling and the details of the two dark matter halo models are explained in Section \ref{sec:rotcurve_decomp}. Our modelling strategy is described in Section \ref{sec:fit_proc} and the results are presented in Section \ref{sec:results}. Finally we list our main conclusions in Section \ref{sec:conclusions}.

\section{Sample selection}
\label{sec:sample}

Our sample of rotation curves was compiled mainly from the Little THINGS \citep{hunter12, iorio16}, THINGS \citep{walter08,deblok08}, and SPARC \citep{lelli16} datasets. For Little THINGS we used the publicly available rotation curve data from \cite{iorio16} and took the surface density and surface brightness profiles of the atomic gas and the stars from \cite{oh15}. The latter were kindly provided to us by S.H. Oh and D. Hunter. The THINGS data, both the rotation curves and the baryonic profiles, were kindly made available by E. de Blok. Finally the SPARC data are publicly available and can be downloaded from the SPARC website\footnote{http://astroweb.cwru.edu/SPARC/}. For all three datasets the stellar surface brightness profiles are based on observations at 3.6 $\mu$m.
\par
For Little THINGS we selected only the galaxies that are marked as `clean dIrrs' by \cite{read16c} and further eliminated NGC\,6822 and DDO\,210. For the former no rotation curve is presented by \cite{iorio16}, while the rotation curve of the latter is highly uncertain and completely dominated by the asymmetric drift correction. This leaves 9 galaxies from the Little THINGS sample. \cite{deblok08} present rotation curves of 19 THINGS galaxies, from which we eliminated 10 because of poor sampling of the rising part of the rotation curve or strong non-circular motions. The gap in mass between the Little THINGS and THINGS galaxies is bridged with a set of low mass systems from the SPARC dataset. These are selected according to the following criteria: a total 3.6 $\mu$m luminosity $L_{3.6}$ $\lesssim$ 10$^9$ $L_{\odot}$, a rotation curve with quality label 1, a reliable distance estimate, an inclination between 40$\degr$ and 80$\degr$, and little beam smearing. These criteria lead to an additional 5 galaxies. Finally we also included the rotation curve of M33 \citep[taken from][]{corbelli14}, which is a galaxy previously claimed to have a strongly cusped dark matter halo \citep{corbelli14,hague15}. For this galaxy \cite{corbelli14} have not reported the stellar surface brightness profile, but have immediately derived the surface \emph{density} profile of the stars from a pixel by pixel population synthesis analysis.
\par
Our total sample thus comprises 24 rotation curves. For three of the THINGS galaxies in this sample, however, the rotation curves might actually be unreliable owing to a substantial bar (NGC\,925), poorly constrained distance (NGC\,3521), or uncertain inclination in the outer half (NGC\,7793). Since \cite{hague14} have included these rotation curves in their analysis, we also kept them in our sample, but marked them as problematic and only showed their fits without including them in the further analysis. UGC\,8490, from the SPARC dataset, was similarly marked problematic because \cite{mcquinn15} have discovered a strong increase in its star formation rate over the past 100 Myr. This indicates that UGC\,8490 might be experiencing a starburst, which may severely bias its kinematics. 
\par
Finally, \cite{iorio16} have remarked that their rotation curves of UGC\,8508 and DDO\,126 are unreliable up to a radius of 0.5 and 1.43 kpc, respectively, while \cite{gentile07} noted that elliptical streaming motions could be affecting the innermost data points of their rotation curve of NGC\,3741, up to a radius of 1.2 kpc. \cite{blais01} additionally found that the inner seven data points of the rotation curve of NGC\,3109 could be slightly underestimated because of weak beam smearing. These same data points also have suspiciously small error bars. Since only the inner parts of the rotation curves are affected, we still included these galaxies in our `good' sample, but excluded the affected data points from the fits. The rotation curves are still sampled well enough by the remaining points.
\par
An overview of our complete sample is given in Table \ref{table:sample}. For the galaxies from the THINGS and SPARC datasets we used distances from the Cosmicflows-2 catalogue \citep{tully13}. For M33 we took over the distance from \cite{corbelli14} and for the Little THINGS galaxies we used the distances from \cite{iorio16}.

\begin{table*}[t]
\caption{Sample overview. Columns 1-3 represent the galaxy name and distance and an indicative inclination. Columns 4-5 give the integrated 3.6 $\mu$m luminosity of the disk and bulge (if present). Column 6 gives the stellar half-light radius. Comments on the quality and data references are given in columns 7-8.}
\label{table:sample}
\centering
\begin{threeparttable}
\begin{tabular}{l c c c c c c c}
\hline \hline\noalign{\smallskip}
\smallskip
Galaxy & D & incl & log$_{10}$\,($L^{\text{D}}_{3.6}$) & log$_{10}$\,($L^{\text{B}}_{3.6}$) & $R_{1/2}$ & Comment & Reference \\
\smallskip
 & (Mpc) & ($\degr$) & ($L_{\odot}$) & ($L_{\odot}$) & (kpc) & & \\
\hline\noalign{\smallskip}
UGC\,8508 & 2.6 & 68 & 6.954 &  & 0.623 &  & 1,2 \\
CVnIdwA & 3.6 & 49 & 7.199 &  & 1.784 &  & 1,2 \\
NGC\,3741 & 3.23 & 70 & 7.453 &  & 0.341 &  & 3 \\
WLM & 1.0 & 74 & 7.522 &  & 1.334 &  & 1,2 \\
DDO\,154 & 3.7 & 68 & 7.628 &  & 3.511 &  & 1,2 \\
DDO\,126 & 4.9 & 62 & 7.880 &  & 1.771 &  & 1,2 \\
DDO\,87 & 7.4 & 43 & 8.121 &  & 3.363 &  & 1,2 \\
UGCA\,442 & 4.37 & 64 & 8.150 &  & 1.906 &  & 3 \\
DDO\,168 & 4.3 & 47 & 8.168 &  & 1.562 &  & 1,2 \\
DDO\,52 & 10.3 & 55 & 8.300 &  & 2.278 &  & 1,2 \\
NGC\,3109 & 1.37 & 70 & 8.314 &  & 2.700 &  & 3 \\
NGC\,2366 & 3.4 & 65 & 8.548 &  & 2.899 &  & 1,2 \\
UGC\,7603 & 6.85 & 78 & 8.902 &  & 1.239 &  & 3 \\
\hline\noalign{\smallskip}
\multicolumn{8}{c}{DC14 only} \\
\hline\noalign{\smallskip}
IC\,2574 & 3.89 & 51 & 9.352 &  & 4.352 &  & 4 \\
NGC\,2976 & 3.63 & 54 & 9.516 &  & 1.505 &  & 4 \\
M33 & 0.84 & 55 & 9.690\tnote{*} &  & 3.325 &  & 5 \\
NGC\,2403 & 3.18 & 55 & 10.074 & 8.846 & 2.215 &  & 4 \\
NGC\,3621 & 6.73 & 62 & 10.536 &  & 4.471 &  & 4 \\
NGC\,3198 & 13.37 & 72 & 10.520 & 9.569 & 5.209 &  & 4 \\
NGC\,5055 & 9.04 & 51 & 11.096 & 10.182 & 4.100 &  & 4 \\
\hline\noalign{\smallskip}
\multicolumn{8}{c}{Problematic galaxies} \\
\hline\noalign{\smallskip}
UGC\,8490 & 4.76 & 50 & 9.028 &  & 1.167 & a & 3 \\
NGC\,7793 & 3.58 & 43 & 9.874 &  & 1.836 & b & 4 \\
NGC\,925 & 8.91 & 50 & 10.168 &  & 11.407 & c & 4 \\
NGC\,3521 & 14.2 & 69 & 11.472 &  & 4.514 & d & 4 \\
\hline \hline\noalign{\smallskip}
\end{tabular}
\end{threeparttable}
\tablebib{
(1) Little THINGS \citep{iorio16}; (2) Little THINGS \citep{oh15}; (3) SPARC \citep{lelli16}; (4) THINGS \citep{deblok08}; (5) \cite{corbelli14}.
\textbf{Comments.} (a) Potential starburst (b) uncertain inclination for outer disk; (c) extended bar; and (d) poorly constrained distance.\\
* For M33 column 4 gives the stellar \emph{mass} from \cite{corbelli14} (in $M_{\odot}$) instead of the 3.6 $\mu$m luminosity.
}
\end{table*}

\section{Rotation curve decomposition and halo models}
\label{sec:rotcurve_decomp}

In a disk galaxy the inward gravitational force that pulls a particle towards the centre is balanced by the outward centripetal acceleration from its rotation. The total gravitational potential acting on this particle is the sum of the potentials from the individual components: gas, stars, and dark matter. We can therefore write
\begin{equation}
a_{\text{cpt}} = a_{\text{grav}} = a_{\text{gas}} + a_{*} + a_{\text{dm}}.
\end{equation}
Since the centripetal acceleration is proportional to the square of the circular velocity, this can be re-written as
\begin{equation}
\label{eq:vs_1}
v^{2}_{\text{c}} = v^{2}_{\text{gas}} + v^{2}_{*} + v^{2}_{\text{dm}},
\end{equation}
where $v_{\text{gas}}$, $v_{*}$, and $v_{\text{dm}}$ are the circular velocities needed to balance the gravitational force exerted by the gas, stars, and dark matter, respectively. These are of course related to the mass distributions of the individual components. For the stars, however, the conversion between the observed luminosity and the mass is uncertain. The mass-to-light ratio $\Upsilon$ is therefore isolated as an unknown parameter in equation \ref{eq:vs_1}. In addition, the distribution of the gas often shows a hole in the centre. Inside this hole the gravitational pull from the gas is directed outward, giving a negative contribution to the total circular velocity. A better formulation of equation \ref{eq:vs_1} is therefore
\begin{equation}
\label{eq:vs_2}
v^{2}_{\text{c}} = v_{\text{gas}}\,|v_{\text{gas}}| + \Upsilon_{*}v^{2}_{*} + v^{2}_{\text{dm}},
\end{equation}
or
\begin{equation}
\label{eq:vs_3}
v^{2}_{\text{c}} = v_{\text{gas}}\,|v_{\text{gas}}| + \Upsilon_{*,\text{B}}\,v^{2}_{*,\text{B}} + \Upsilon_{*,\text{D}}\,v^{2}_{*,\text{D}} + v^{2}_{\text{dm}}
\end{equation}
if the stellar distribution is decomposed in bulge and disk components. The value $v_{*}$ is now the circular velocity from the stars for a mass-to-light ratio of 1.
\par
Equations \ref{eq:vs_2} and \ref{eq:vs_3} form the basis for the mass modelling performed in this work. The total circular velocity $v_{\text{c}}$ is measured by the rotation curve (although, see section \ref{sec:ad_correction}), while the gas and stellar circular velocities $v_{\text{gas}}$ and $v_{*}$ are derived from their observed surface brightness profiles. For this purpose a thin disk geometry is generally assumed for the gas. The stellar distribution is usually modelled as a thick disk with an exponential or sech$^2$ profile in the vertical direction. This leaves the stellar mass-to-light ratio(s) and dark matter contribution $v^{2}_{\text{dm}}$ as the only unknowns. For the latter we use two different parameterizations: the \textsc{core}NFW halo and DC14 halo. We express both these parameterizations in terms of the virial radius and virial mass. The former is defined as the radius inside which the average density of the dark matter halo is equal to $\Delta$ times the critical density of the Universe $\rho_{\text{crit}}$, where $\Delta$ and $\rho_{\text{crit}}$ depend on the assumed cosmology. The virial mass is simply the enclosed mass at the virial radius,
\begin{equation}
M_{\text{vir}} = \frac{4}{3} \pi \, r^{3}_{\text{vir}} \, \Delta \, \rho_{\text{crit}}.
\end{equation} 
For consistency with \cite{dicintio14b} we use a WMAP3 cosmology \citep{spergel07} with $\Delta$ = 93.6, $H_0$ = 73.0 km s$^{-1}$ Mpc$^{-1}$ and $\rho_{\text{crit}}$ = 147.896 $M_{\odot}$ kpc$^{-3}$.

\subsection{DC14}

The DC14 profile is formulated by \cite{dicintio14b} as a special case of the general and very flexible ($\alpha$,\,$\beta$,\,$\gamma$) profile \citep{jaffe83, hernquist90, zhao96}
\begin{equation}
\label{eq:abg}
\rho(r) = \frac{\rho_{\text{s}}}{\Big(\frac{r}{r_{\text{s}}}\Big)^{\gamma} \Big[1+\Big(\frac{r}{r_{\text{s}}}\Big)^{\alpha}\Big]^{(\beta-\gamma)/\alpha}}.
\end{equation}
At small and large radii this profile follows a power law with slopes $\gamma$ and $\beta$, respectively, and the sharpness of the transition between these two regimes is governed by $\alpha$. This profile reduces to a simple Navarro-Frenk-White (NFW) profile for ($\alpha$,\,$\beta$,\,$\gamma$) = (1,\,3,\,1) and the frequently used pseudo-isothermal halo is recovered when ($\alpha$,\,$\beta$,\,$\gamma$) = (2,\,2,\,0). \cite{hague14,hague15} have recently used this profile in its most general form to model the dark matter haloes of M33 and a sample of THINGS galaxies.
\par
\begin{figure}
\centering
\resizebox{\hsize}{!}{\includegraphics{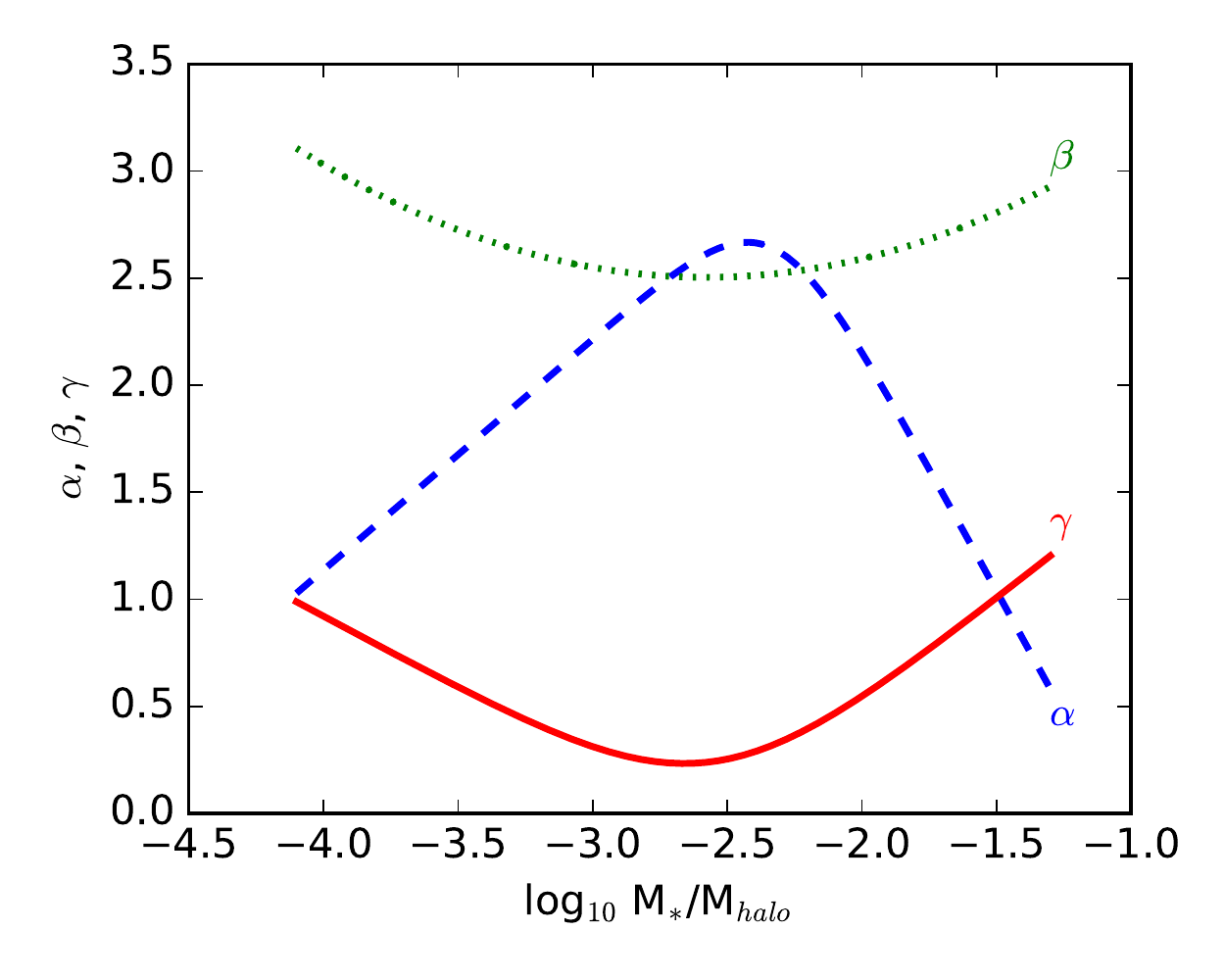}}
\caption{Variation of the shape parameters $\alpha$, $\beta,$ and $\gamma$ of the DC14 halo as a function of the integrated star formation efficiency.}
\label{fig:dc14_shape_params}
\end{figure}

\smallskip
Starting from an NFW profile, but in the general formulation of equation \ref{eq:abg}, \cite{dicintio14b} allow the modification of the inner slope by stellar feedback by expressing the shape parameters $\alpha$, $\beta,$ and $\gamma$ as a function of the integrated star formation efficiency $M_{*}/M_{\text{halo}}$ as follows:
\begin{equation}
\begin{aligned}
&\alpha = 2.94 - \textrm{log}_{10}[(10^{X+2.33})^{-1.08} + (10^{X+2.33})^{2.29}] \\
&\beta = 4.23 + 1.34X + 0.26X^2 \\
&\gamma = -0.06 + \textrm{log}_{10}[(10^{X+2.56})^{-0.68} + (10^{X+2.56})],
\end{aligned}
\end{equation}
with $X$ = log$_{10}$\,($M_{*}/M_{\text{halo}}$). These expressions are only valid for $-4.1 < X < -1.3$, which is the range probed by the simulations of \cite{dicintio14b}. At lower values of $X$, too few stars form to modify the dark matter halo. On the other hand, at $X$ > -1.3 (corresponding to halo masses $\gtrsim$ 10$^{12}$ $M_{\odot}$) processes not included in the simulations, such as AGN feedback, can start to play a role as well.\\
The variation of $\alpha$, $\beta,$ and $\gamma$ as a function of $X$ is shown in Fig. \ref{fig:dc14_shape_params}. The inner log slope $\gamma$ first decreases with increasing $X$, since a higher stellar-to-halo mass ratio implies more energy input from supernova feedback. However, it reaches a minimum at $X \sim -2.6$ and goes back up at higher values of $X$. The reason for this turnover is the increasing gravitational potential of the stars, which at a certain point starts to dominate the feedback and pulls the dark matter back towards the centre. In the DC14 formalism more star formation therefore does not monotonically result in ever stronger cores.

\par
\smallskip
For an NFW halo the concentration is defined as $c$ = $r_{\text{vir}}/r_{\text{s}}$, where the scale radius $r_{\text{s}}$ is equal to $r_{-2}$, the radius at which the slope of the density profile becomes -2. For the ($\alpha$,\,$\beta$,\,$\gamma$) profile, the meaning of $r_{\text{s}}$ depends on the values of $\alpha$, $\beta,$ and $\gamma$, with
\begin{equation}
r_{-2} = \Bigg(\frac{2-\gamma}{\beta-2}\Bigg)^{1/\alpha}\,r_{\text{s}}.
\end{equation}
\cite{dicintio14b} therefore define the concentration of their dark matter haloes as
\begin{equation}
c_{\text{vir}} = \frac{r_{\text{vir}}}{r_{-2}}.
\end{equation}
The concentration of the original, unmodified NFW halo can be recovered from this as
\begin{equation}
c_{\text{NFW}} = \frac{c_{\text{vir}}}{1.0+0.00003e^{3.4(X+4.5)}},
\end{equation}
with again $X$ = log$_{10}$\,($M_{*}/M_{\text{halo}}$). It is this concentration that should be used to compare the DC14 halo from a fit to a rotation curve to, for example  the mass-concentration relation.

\subsection{coreNFW}

A \textsc{core}NFW \citep{read16a} halo is essentially a NFW halo with the inner part modified by a spherically symmetric function $f^n$ that models the effects of supernova feedback. Practically this modification is expressed at the level of the enclosed mass. For an ordinary NFW halo profile \citep{navarro96}
\begin{equation}
\rho_{\text{NFW}}(r) = \frac{\rho_{\text{s}}}{\Big(\frac{r}{r_{\text{s}}}\Big) \Big(1+\frac{r}{r_{\text{s}}}\Big)^{2}}
\end{equation}
with concentration
\begin{equation}
c = r_{\text{vir}} / r_{\text{s}}
\end{equation}
the enclosed mass at a radius $r$ is given by
\begin{equation}
\begin{split}
M_{\text{NFW}}(<r) & = M_{\text{vir}} \, \frac{\textrm{ln}\,(1+r/r_{\text{s}}) - (r/r_{\text{s}})/(1+r/r_{\text{s}})}{\textrm{ln}\,(1+c) - c/(1+c)} \\
            & = M_{\text{vir}} \, g_c \, \Bigg[\textrm{ln}\,\bigg(1+\frac{r}{r_{\text{s}}}\bigg) - \bigg(\frac{r}{r_{\text{s}}}\bigg)\,\bigg(1+\frac{r}{r_{\text{s}}}\bigg)^{-1}\Bigg].
\end{split}
\end{equation}
The \textsc{core}NFW profile is then defined as
\begin{equation}
M_{\text{cNFW}}(<r) = M_{\text{NFW}}(<r)\,f^{n}(r),
\end{equation}
with
\begin{equation}
f(r) = \Bigg[\textrm{tanh} \, \Bigg(\frac{r}{r_{\text{c}}}\Bigg)\Bigg].
\end{equation}
The radial extent of the core is determined by the core radius $r_c$, which \cite{read16a} relate to the stellar half-mass radius as $r_{\text{c}}$ = $\eta \, r_{1/2}$, with an optimal value of 1.75 for the fitting parameter $\eta$ .\\
The strength of the core is governed by the parameter $n$, which ranges between 0 < $n$ $\leq$ 1 and is defined as
\begin{equation}
n =     \textrm{tanh} \, \Bigg(\kappa \frac{t_{\text{SF}}}{t_{\text{dyn}}}\Bigg).
\end{equation}
Here $\kappa$ is again a fitting parameter and the star formation time $t_{\text{SF}}$ is the total time that the galaxy has been forming stars. The dynamical time $t_{\text{dyn}}$ is the duration of 1 circular orbit at the scale radius in the unmodified NFW halo
\begin{equation}
t_{\text{dyn}} = \frac{2 \pi \, r_{\text{s}}}{v_{\text{NFW}}(r_{\text{s}})} = 2 \pi \sqrt{\frac{r^{3}_s}{G M_{\text{NFW}}(<r_{\text{s}})}}.
\end{equation}
The longer stars have been forming, the larger $n$ and stronger the core. On the other hand, the bigger the original dark matter halo, the smaller $n$ and more difficult it is to form a core. Following \cite{read16a}, we set $\kappa$ = 0.04 and choose $t_{\text{SF}}$ = 14 Gyrs.

\subsection{Asymmetric drift correction}
\label{sec:ad_correction}

The gravitational attraction from the gas, stars, and dark matter is in fact not balanced solely by circular motion, but also for a small part by the internal pressure of the gas. The observed rotation velocity $v_{\text{rot}}$ is therefore not exactly equal to the circular velocity $v_{\text{c}}$ from equation \ref{eq:vs_2}. Instead it is given by
\begin{equation}
v^{2}_{\text{rot}} = v^{2}_{\text{c}} + \Bigg[\frac{R}{\rho} \frac{\partial (\rho\sigma^{2}_{\text{R}})}{\partial R} + \sigma^{2}_{\text{R}} - \sigma^{2}_{\phi} + R\frac{\partial (\overline{v_{\text{R}}v_{\text{z}}})}{\partial z}\Bigg]
\end{equation}
\citep[equation 4-227 of][]{binney08}, where $\rho$ and $\sigma$ are the density and velocity dispersion of the gas. The asymmetric drift correction (term inside the square brackets) is usually simplified under the assumptions that the velocity dispersion is isotropic ($\sigma_{\text{R}}$ = $\sigma_{\phi}$), the velocity ellipsoid is aligned with the cylindrical coordinate system ($\overline{v_{\text{R}}v_{\text{z}}}$ = 0), and the vertical scale height does not change much with radius. This leads to
\begin{equation} \label{eq:correction}
v^{2}_{\text{c}} = v^{2}_{\text{rot}} - \frac{R}{\Sigma} \frac{\partial (\Sigma\sigma^{2})}{\partial R},
\end{equation}
where $\Sigma$ is the surface density of the gas. The observationally derived radial $\Sigma\sigma^{2}$ profile is typically rather rugged, leading to sometimes strong and unphysical fluctuations in its derivative. To avoid this, a smooth function is fitted to the profile and the derivative is determined analytically. The simplifications involved in deriving equation \ref{eq:correction} limit its accuracy. As a consequence, equation \ref{eq:correction} only provides an order of magnitude estimate of the correction.
\par

\begin{figure}[h]
\centering
\resizebox{\hsize}{!}{\includegraphics{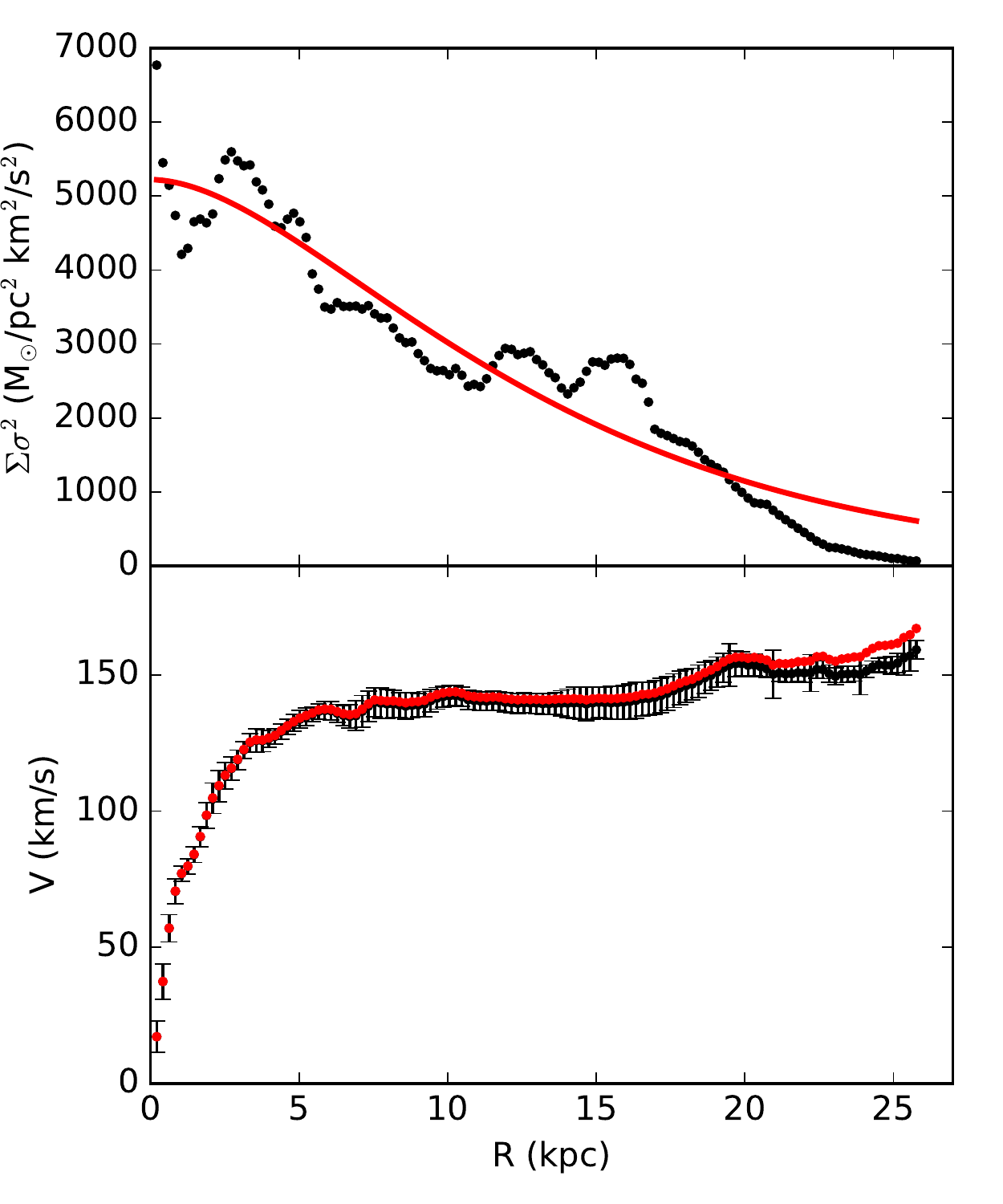}}
\caption{Asymmetric drift correction for NGC\,3621. Top: observed $\Sigma\sigma^{2}$ profile (black circles) and analytic fit (red line) are shown. Bottom: original (black circles and error bars) and corrected (red circles) rotation curves are shown.}
\label{fig:drift_correction}
\end{figure}

For the rotation curves taken from the Little THINGS and SPARC datasets the asymmetric drift correction is already taken into account by the authors. On the other hand, \cite{deblok08} and \cite{corbelli14} did not consider asymmetric drift for their THINGS and M33 rotation curves. We therefore evaluated this correction based on equation \ref{eq:correction}. Depending on the shape of the $\Sigma\sigma^{2}$ profile, we used one of the following analytic functions:
\begin{equation}
\Sigma\sigma^{2}(R) = I_{0} \frac{R_{0}+1}{R_{0}+e^{\alpha R}}
\end{equation}
for a profile with a central core \citep{oh11}, and
\begin{equation}
\Sigma\sigma^{2}(R) = I_{0} \Bigg(1+\frac{R}{R_0}\Bigg)^{\alpha} e^{-\frac{R}{R_0}}
\end{equation}  
for a profile showing a hole in the centre \citep{read16b}. In the inner halves of the rotation curves the derived corrections are consistently much smaller than the error bars and generally only of the order of 1 km s$^{-1}$ or less. For four galaxies the corrections become more substantial (of order $5 - 10$ km s$^{-1}$) near the outer edge of the rotation curve. However, because these larger corrections occur far from the centre, we found that they have only little effect on our fits and do not change any of our conclusions. In addition, the agreement between the $\Sigma\sigma^{2}$ profile and the analytic function is often not very good in these regions, making the corrections uncertain. This is illustrated for NGC\,3621 in Fig. \ref{fig:drift_correction}. Because of this uncertainty and the very limited effect the corrections have on our results, we decided to use the original, uncorrected rotation curves in our analysis.

\section{Markov Chain Monte Carlo fitting}
\label{sec:fit_proc}

\subsection{\texttt{emcee}}

The dynamical models were fitted to the rotation curves with \texttt{emcee}\footnote{http://dan.iel.fm/emcee/current/} \citep{foreman-mackey13}, which is an open-source python implementation of the affine invariant MCMC ensemble sampler from \cite{goodman10}. Markov Chain Monte Carlo or MCMC \citep[e.g.][]{metropolis53,hastings70,press07} is a sampling technique that has been applied to the decomposition of rotation curves for several years \citep[e.g.][]{puglielli10,hague13}. It is more efficient in sampling the parameter space than the fitting techniques used in earlier works and has the big advantage that it returns the full multidimensional probability distribution of all the parameters instead of only the best-fit model. In addition MCMC allows us to include physical knowledge about the parameters in the fits via so-called priors that are combined with the likelihood function.
\par
The \texttt{emcee} algorithm explores the N-dimensional parameter space with different, randomly initialized walkers that each make their own MCMC chain. An initial burn-in phase is used to allow the walkers to move to the relevant high-likelihood areas of the parameter space. After this the walkers are reinitialized at their current positions and the actual MCMC chains are made. As a last step the chains of all the walkers are combined to form the final MCMC chain. For every fit we used 100 walkers, each taking 2000 steps of which the first 1000 were used as burn-in. These numbers are in line with the \texttt{emcee} recommendations \citep{foreman-mackey13} and the values typically used in other works \citep[e.g.][]{kirichenko15, katz16, read16c}, and ensured good convergence of our fits (see below). Our likelihood function is
\begin{equation}
\mathcal{L} = e^{-{\chi}^{2}/2}.
\end{equation}
\par
For a multi-modal posterior distribution, part of the walkers can get stuck in isolated low probability modes if they are initialized randomly over the full range of the parameter space (within the imposed boundaries). This generates numerous irrelevant peaks in the retrieved posterior distribution. We therefore performed each fit in two iterations. First the walkers were initialized randomly over the full relevant range of parameter space. The different peaks in the posterior distribution were then investigated to find the mode with the highest likelihood. Next, as a second iteration, we redid the fit with the walkers now initialized in a small Gaussian ball centred on this mode and with $\upsigma$ equal to 1 percent of the allowed range for each parameter. The parameter values that are used in the figures below and reported in Table \ref{table:results} correspond to the maximum likelihood model for each fit.\\
For good performance, an MCMC sampler should be run for at least a few (about 10) autocorrelation times and should have an acceptance fraction between 0.2 and 0.5 \citep{foreman-mackey13}. With 1000 steps taken by each walker, the first condition was well met for all the fits. Appropriately setting  the \texttt{emcee} proposal scale parameter to a value of 2 or 3 ensured that the second condition was also met. Finally we checked the convergence of the MCMC chains by performing each fit three times and evaluating the Gelman-Rubin eigenvalues with the \texttt{GetDist}\footnote{https://pypi.python.org/pypi/GetDist/} python package. These values were well below 1 for all the fits, indicating good convergence.

\subsection{Priors and parameter ranges}

The \textsc{core}NFW halo fits were performed with log$_{10}$\,$M_{\text{vir}}$, $c,$ and $\Upsilon$ (or $\Upsilon_{\text{d}}$ and $\Upsilon_{\text{b}}$) as free parameters. We use the log of $M_{\text{vir}}$ instead of $M_{\text{vir}}$ itself as a parameter in the fits because of the large dynamical range involved. Flat priors were assumed for all free parameters. Log$_{10}$\,$M_{\text{vir}}$ and $c$ were loosely constrained inside 8 < log$_{10}$\,($M_{\text{vir}}/M_{\odot}$) < 14 and 1 < $c$ < 100. The 3.6 $\mu$m mass-to-light ratio was confined to the range 0.3 < $\Upsilon_{3.6}$ < 0.8, as motivated by the constraints from \cite{meidt14} and \cite{mcgaughschombert14}. For M33 we allow the initial stellar mass to vary by a factor 0.758 < $\Upsilon$ < 1.319 based on the uncertainty that is mentioned in Section 6 of \cite{corbelli14}. Following \cite{read16c}, $\eta$, $\kappa,$ and $t_{\text{SF}}$ were kept fixed at 1.75, 0.04, and 14\,Gyrs, respectively. 
\par
For the fits with the DC14 halo, we let $V_{\text{vir}}$, $c_{\text{vir}}$, and $\Upsilon$ (or $\Upsilon_{\text{d}}$ and $\Upsilon_{\text{b}}$) free and again used a flat prior for each of these. Following \cite{katz16} we used wide ranges of 10 < $V_{\text{vir}}/(km\,s^{-1})$ < 500 and 1 < $c_{\text{vir}}$ < 100 for the first two parameters and the same range as before for the mass-to-light ratio: 0.3 < $\Upsilon_{3.6}$ < 0.8 (and 0.758 < $\Upsilon$ < 1.319 for M33).
\par
Since our goal is to find models that fit the rotation curves well and yield physically acceptable dark matter haloes at the same time, we further imposed the cosmological halo mass-concentration and stellar mass-halo mass relations as log-normal priors in the fits. For the DC14 halo the shape parameters $\alpha$, $\beta,$ and $\gamma$ are expressed as a function of log$_{10}$\,($M_{*}/M_{\text{halo}}$), where $M_{\text{halo}}$ = $M_{\text{vir}}$. Hence, $\alpha$, $\beta,$ and $\gamma$ depend on the definition of the virial mass, which depends on the assumed cosmology. Since \cite{dicintio14b} have assumed a WMAP3 cosmology, we did the same in our fits and we used the $M_{\text{halo}}$-$c$ relation from \cite{maccio08} that was derived under this cosmology. The $M_{*}$-$M_{\text{halo}}$ relation from \cite{moster10} has also used the WMAP3 values. However, this relation was derived from abundance matching using the SDSS DR3 stellar mass function for halo masses down to $\sim 3 \times 10^{10}$ $M_{\odot}$ and is an extrapolation at lower masses. As pointed out by \cite{read16c}, this extrapolation is not consistent with the newer and deeper SDSS data, which means that the \cite{moster10} relation is actually not reliable for halo masses below $\sim 3 \times 10^{10}$ $M_{\odot}$. Indeed the stellar mass-halo mass relation from \cite{behroozi13}, which is based on the newer SDSS data and the cosmological parameters used in the Bolshoi simulations \citep[compatible with WMAP5 and WMAP7;][]{klypin11}, is much shallower and diverges significantly from the \cite{moster10} relation at low halo masses. Since a number of the galaxies in our sample fall in this low mass regime we opted to use the relation from \cite{behroozi13} rather than that from \cite{moster10} as a prior in our fits. For \textsc{core}NFW the parameterization is independent of the assumed cosmology and depends only on the global original NFW profile (i.e. before alteration by stellar feedback) and on the stellar half-mass radius and total star formation time.
\begin{table*}[t!]
\caption{Parameters and fit quality of the maximum likelihood DC14 and \textsc{core}NFW models for the good galaxies in our sample. Column 1 gives the galaxy name. Columns 2-5 represent the halo mass, halo concentration, stellar mass, and reduced chi-squared of the best-fit DC14 model. Columns 6-10 give the halo mass, halo concentration, stellar mass, dark matter core radius, and reduced chi-squared of the best-fit \textsc{core}NFW model.}
\label{table:results}
\centering
\begin{threeparttable}
\begin{tabular}{lccccccccc}
\hline \hline \\[-0.2cm]
 & \multicolumn{4}{c}{DC14} & \multicolumn{5}{c}{\textsc{c}NFW} \\
\cmidrule(lr){2-5}
\cmidrule(lr){6-10}
Galaxy & log$_{10}$\,($M_{\text{vir}}$) & c & log$_{10}$\,($M_{*}$) & $\chi^{2}_{\text{red}}$ & log$_{10}$\,($M_{\text{vir}}$) & c & log$_{10}$\,($M_{*}$) & $r_{\text{c}}$ & $\chi^{2}_{\text{red}}$ \\[0.1cm]
 & ($M_{\odot}$) &  & ($M_{\odot}$) &  & ($M_{\odot}$) &  & ($M_{\odot}$) & (kpc) & \\[0.1cm]
\hline \\[-0.2cm]
UGC\,8508 & 9.83$^{+0.22}_{-0.30}$ & 30.84$^{+6.60}_{-5.49}$ & 6.86$^{+0.00}_{-0.42}$ & 0.42 & 9.85$^{+0.26}_{-0.24}$ & 25.43$^{+9.23}_{-6.61}$ & 6.86$^{+0.00}_{-0.27}$ & 1.09 & 0.61 \\[0.2cm]
CVnIdwA & 9.57$^{+0.44}_{-0.24}$ & 16.84$^{+3.21}_{-2.97}$ & 6.68$^{+0.42}_{-0.00}$ & 0.34 & 9.84$^{+0.21}_{-0.29}$ & 13.39$^{+6.97}_{-4.02}$ & 7.04$^{+0.06}_{-0.37}$ & 3.12 & 0.41 \\[0.2cm]
NGC\,3741 & 10.43$^{+0.12}_{-0.10}$ & 15.52$^{+2.02}_{-1.91}$ & 7.35$^{+0.00}_{-0.16}$ & 0.18 & 10.55$^{+0.18}_{-0.16}$ & 9.88$^{+1.92}_{-1.66}$ & 7.35$^{+0.00}_{-0.14}$ & 0.60 & 1.23 \\[0.2cm]
NGC\,3741\tnote{*} &  &  &  &  & 10.35$^{+0.20}_{-0.18}$ & 16.75$^{+9.59}_{-5.22}$ & 7.35$^{+0.00}_{-0.24}$ & 4.06$^{+1.75}_{-1.48}$ & 0.19 \\[0.2cm]
WLM & 10.07$^{+0.32}_{-0.21}$ & 20.94$^{+4.09}_{-5.21}$ & 7.00$^{+0.26}_{-0.00}$ & 0.73 & 10.39$^{+0.24}_{-0.26}$ & 13.07$^{+5.23}_{-2.93}$ & 7.43$^{+0.00}_{-0.17}$ & 2.33 & 0.62 \\[0.2cm]
DDO\,154 & 10.30$^{+0.06}_{-0.06}$ & 20.29$^{+1.36}_{-1.65}$ & 7.53$^{+0.00}_{-0.33}$ & 0.61 & 10.14$^{+0.06}_{-0.06}$ & 46.23$^{+8.30}_{-6.39}$ & 7.53$^{+0.00}_{-0.37}$ & 6.14 & 0.86 \\[0.2cm]
DDO\,126 & 10.04$^{+0.29}_{-0.17}$ & 19.50$^{+3.59}_{-3.69}$ & 7.36$^{+0.42}_{-0.00}$ & 0.17 & 10.33$^{+0.21}_{-0.29}$ & 11.91$^{+7.54}_{-2.63}$ & 7.78$^{+0.00}_{-0.42}$ & 3.1 & 0.40 \\[0.2cm]
DDO\,87 & 10.30$^{+0.14}_{-0.12}$ & 21.74$^{+2.78}_{-3.23}$ & 7.71$^{+0.31}_{-0.12}$ & 0.24 & 10.74$^{+0.19}_{-0.26}$ & 13.96$^{+9.13}_{-3.58}$ & 8.02$^{+0.00}_{-0.23}$ & 5.88 & 1.24 \\[0.2cm]
UGCA\,442 & 10.53$^{+0.07}_{-0.06}$ & 18.99$^{+2.06}_{-1.61}$ & 8.05$^{+0.00}_{-0.39}$ & 0.65 & 10.59$^{+0.17}_{-0.18}$ & 14.00$^{+5.77}_{-3.17}$ & 8.05$^{+0.00}_{-0.36}$ & 3.34 & 0.63 \\[0.2cm]
DDO\,168 & 10.66$^{+0.18}_{-0.18}$ & 18.89$^{+2.32}_{-1.89}$ & 8.03$^{+0.04}_{-0.31}$ & 1.61 & 10.51$^{+0.22}_{-0.19}$ & 16.51$^{+5.25}_{-4.23}$ & 7.81$^{+0.26}_{-0.17}$ & 2.73 & 2.2 \\[0.2cm]
DDO\,52 & 10.32$^{+0.25}_{-0.14}$ & 20.07$^{+2.41}_{-5.11}$ & 7.78$^{+0.42}_{-0.00}$ & 0.20 & 10.64$^{+0.21}_{-0.26}$ & 11.42$^{+6.35}_{-2.69}$ & 8.20$^{+0.00}_{-0.42}$ & 3.99 & 0.30 \\[0.2cm]
NGC\,3109 & 10.86$^{+0.13}_{-0.13}$ & 16.96$^{+1.86}_{-1.50}$ & 8.22$^{+0.00}_{-0.22}$ & 0.13 & 10.83$^{+0.16}_{-0.17}$ & 14.57$^{+4.95}_{-2.98}$ & 8.22$^{+0.00}_{-0.27}$ & 4.73 & 0.23 \\[0.2cm]
NGC\,2366 & 10.57$^{+0.16}_{-0.12}$ & 17.00$^{+1.62}_{-2.50}$ & 8.02$^{+0.33}_{-0.00}$ & 0.93 & 10.57$^{+0.22}_{-0.16}$ & 16.03$^{+6.12}_{-4.98}$ & 8.03$^{+0.32}_{-0.00}$ & 5.07 & 1.25 \\[0.2cm]
UGC\,7603 & 10.60$^{+0.32}_{-0.12}$ & 19.48$^{+4.16}_{-6.68}$ & 8.50$^{+0.31}_{-0.12}$ & 0.43 & 10.81$^{+0.22}_{-0.25}$ & 13.06$^{+5.93}_{-3.37}$ & 8.53$^{+0.19}_{-0.15}$ & 2.17 & 0.58 \\[0.2cm]
\hline \\[-0.2cm]
IC\,2574 & 11.15$^{+0.14}_{-0.09}$ & 9.63$^{+0.29}_{-1.13}$ & 8.83$^{+0.17}_{-0.00}$ & 0.20 &  &  &  &  &  \\[0.2cm]
NGC\,2976 & 11.11$^{+0.16}_{-0.11}$ & 23.69$^{+1.99}_{-3.97}$ & 8.99$^{+0.08}_{-0.00}$ & 0.43 &  &  &  &  &  \\[0.2cm]
M33 & 11.48$^{+0.03}_{-0.03}$ & 9.53$^{+0.50}_{-0.24}$ & 9.81$^{+0.00}_{-0.02}$ & 1.47 &  &  &  &  &  \\[0.2cm]
NGC\,2403 & 11.61$^{+0.04}_{-0.04}$ & 12.57$^{+1.41}_{-0.93}$ & 9.83$^{+0.04}_{-0.05}$ & 0.56 &  &  &  &  &  \\[0.2cm]
NGC\,3621 & 11.85$^{+0.06}_{-0.06}$ & 6.66$^{+0.69}_{-0.50}$ & 10.23$^{+0.03}_{-0.04}$ & 0.56 &  &  &  &  &  \\[0.2cm]
NGC\,3198 & 11.88$^{+0.03}_{-0.07}$ & 4.00$^{+1.31}_{-0.14}$ & 10.44$^{+0.00}_{-0.08}$ & 1.04 &  &  &  &  &  \\[0.2cm]
NGC\,5055 & 11.96$^{+0.05}_{-0.02}$ & 9.59$^{+0.58}_{-2.06}$ & 10.63$^{+0.05}_{-0.00}$ & 0.68 &  &  &  &  &  \\[0.2cm]
\hline \hline\noalign{\smallskip}
\end{tabular}
\begin{tablenotes}
\item[*] \textsc{core}NFW fit with the core radius $r_{\text{c}}$ as a free parameter.
\end{tablenotes}
\end{threeparttable}
\end{table*}

\subsection{Uncertainties}
\label{sec:errorbars}

\begin{figure}[]
\centering
\includegraphics[scale=0.55]{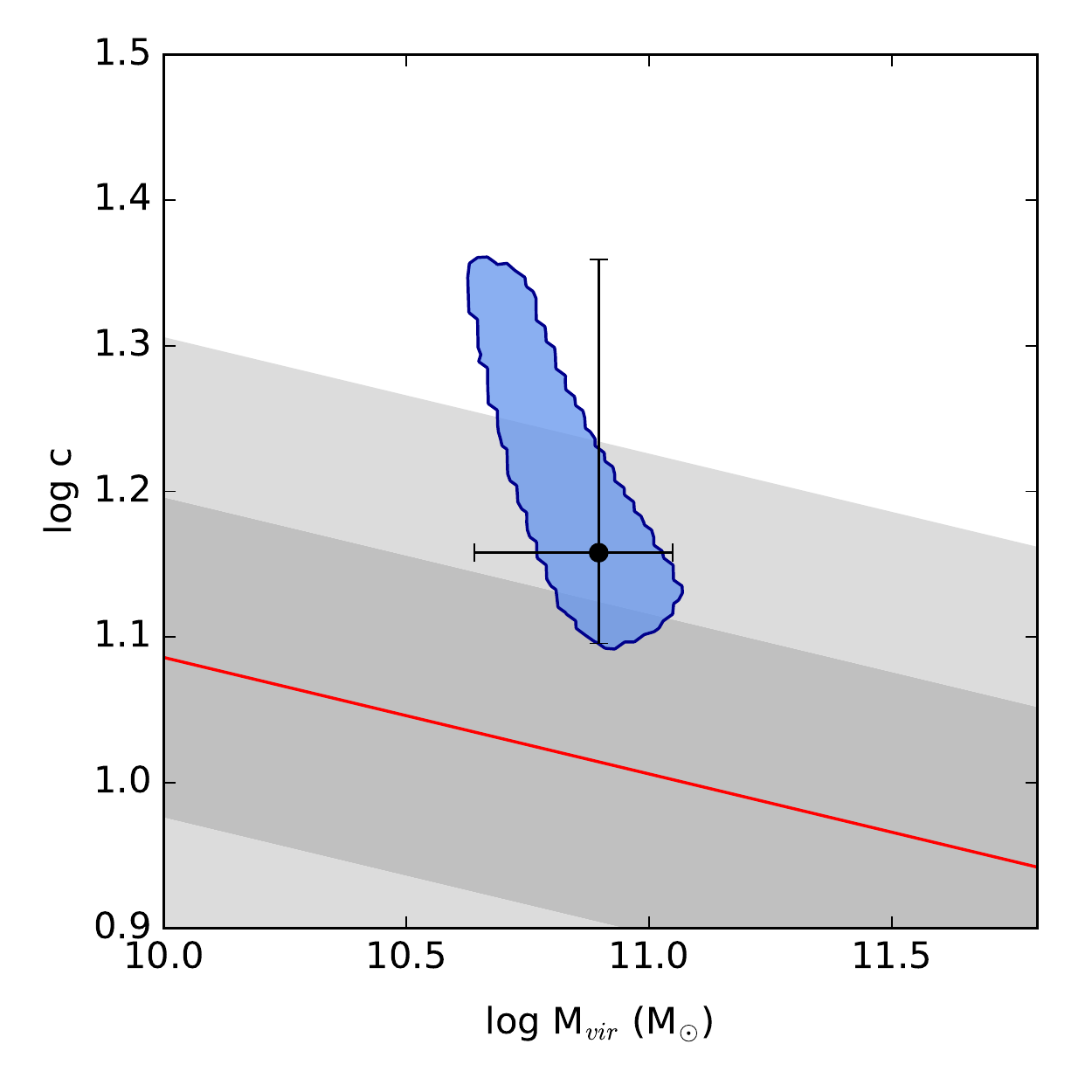}
\caption{Halo mass and concentration of the best-fit DC14 model of UGC\,7603. The projected 68\% confidence region, shown as the blue shaded area, is significantly smaller than the area suggested by the two error bars. Also shown in the plot is the theoretical $M_{\text{halo}}-c$ relation from \cite{maccio08} (red line) and its 1$\sigma$ and 2$\sigma$ scatter (dark and light grey bands).}
\label{fig:contours_vs_errorbars}
\end{figure}

Using the GetDist package, the uncertainties for the different parameters were determined from the multidimensional 68\% confidence region of the full posterior distribution, as the extremal values of the projection of that region onto each parameter axis. As such the error bars give a good indication of how tight the constraints are for a certain parameter, but they should not be over-interpreted as the absolute range of good models. Indeed, if the fit quality of the best-fit model is very high, many models outside of the N-dimensional 68\% confidence region often still provide an acceptable fit to the data. On the other hand, if we plot, for example, the halo mass versus its concentration, the area suggested by the two (orthogonal) error bars is often larger than the actual area to which the models from the MCMC chain are confined (i.e. the projection of the multidimensional confidence region onto the $M_{\text{halo}}$-$c$ plane). This is illustrated in Figure \ref{fig:contours_vs_errorbars} for the best-fit DC14 model of UGC\,7603.

\subsection{$\chi^{2}_{\text{red}}$ and fit quality}

In the discussion of our results we express the quality of the fits to the rotation curves in terms of the reduced chi-squared statistic ($\chi^{2}_{\text{red}}$). In the ideal case where the uncertainties on all the rotation curves are Gaussian and derived in a uniform way, and where all the points of a rotation curve have equal importance, this would be a good measure to compare the fit qualities for all the galaxies in our sample. In reality, however, our rotation curves are compiled from the literature with differing data quality and techniques used to estimate the error bars. In addition some rotation curves keep rising up to the last point, whereas others, for the more massive galaxies, include a large flat part. The latter is generally easier to reproduce and can have a large impact on the $\chi^{2}_{\text{red}}$ value of a fit, but is at the same time much less important in the analysis of core formation. For these reasons the $\chi^{2}_{\text{red}}$ values of our fits are only meaningful to compare the quality of different fits for the same galaxy and not to compare fits for different galaxies.

\section{Results}
\label{sec:results}

Following the procedure outlined in section \ref{sec:fit_proc} we have fitted DC14 halo models to each of the rotation curves in our sample and \textsc{core}NFW models to the rotation curves of the Little THINGS and SPARC galaxies. The best-fit parameters and $\chi^{2}_{\text{red}}$ values of these fits are listed in Table \ref{table:results}. The results of the fits are discussed in the sections below.

\subsection{DC14}
\label{sec:dc14_results}

Figure \ref{fig:dc14_fits} shows the individual DC14 models for the galaxies in our `good' sample. As can be seen, the DC14 halo generally provides excellent fits to the rotation curves, confirming the recent results from both \cite{katz16} and \cite{pace16}. The only clear exception to this is the rotation curve of DDO\,168, where the model overestimates the data in the inner part. However, the inner three points of the rotation curve are in fact already well accounted for by the gravitational potential of the gas alone, so any model with a non-zero contribution of the dark matter at these radii will overestimate the data. Similar arguments also hold for the very inner regions of NGC\,2366 and NGC\,3198. 
\begin{figure*}[t!]
\centering
\includegraphics[width=\textwidth]{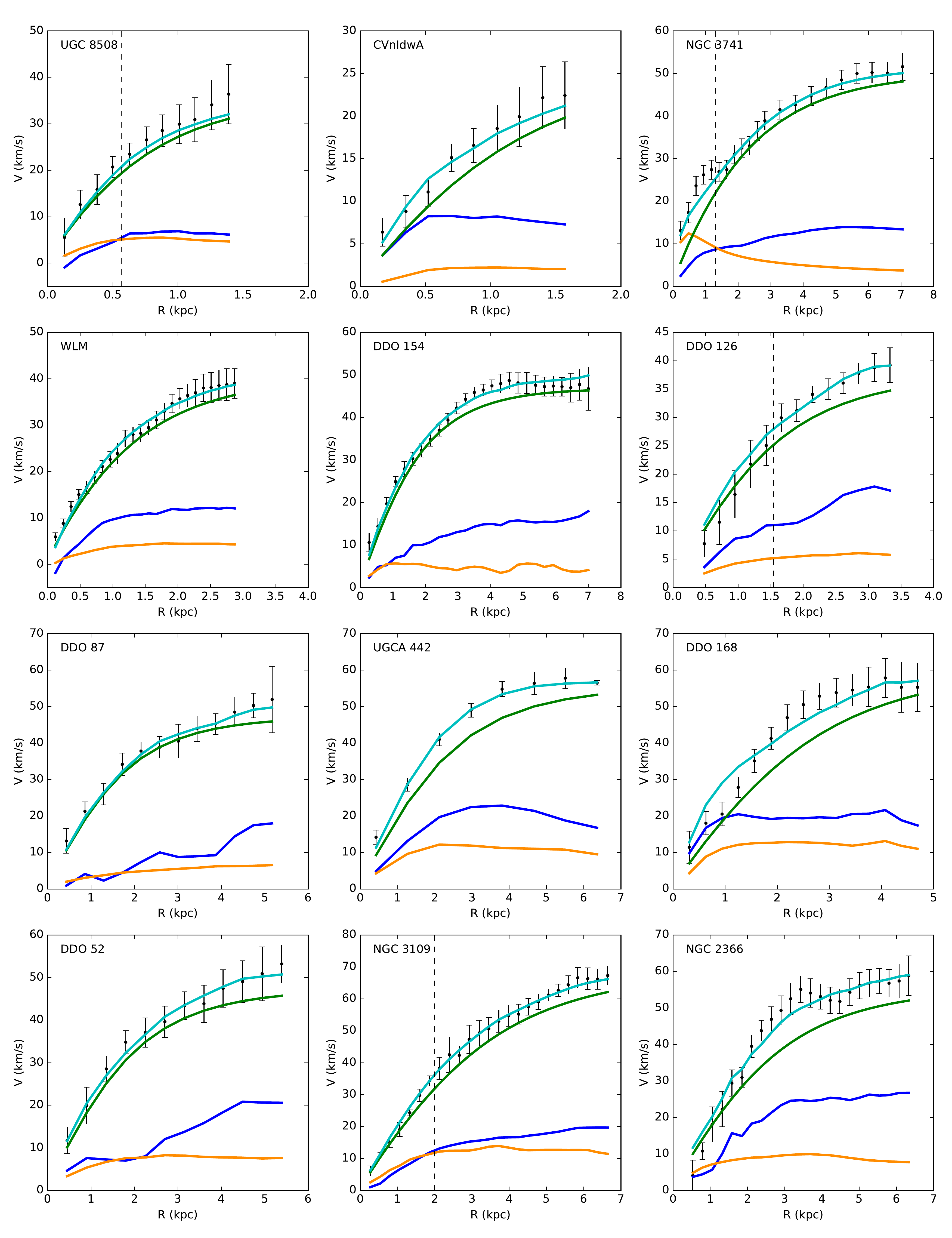}
\caption{Decomposition of the good rotation curves in our sample according to the maximum likelihood DC14 models. The black points show the observed rotation curves and the cyan curves represent the models. The contributions from the individual components are given by the blue (gas), yellow and red (stars), and green (dark matter) curves. The vertical dashed lines indicate the central ranges that were, in some cases, excluded from the fit.}
\label{fig:dc14_fits}
\end{figure*}
\begin{figure*}[]
\ContinuedFloat
\centering
\includegraphics[width=\textwidth]{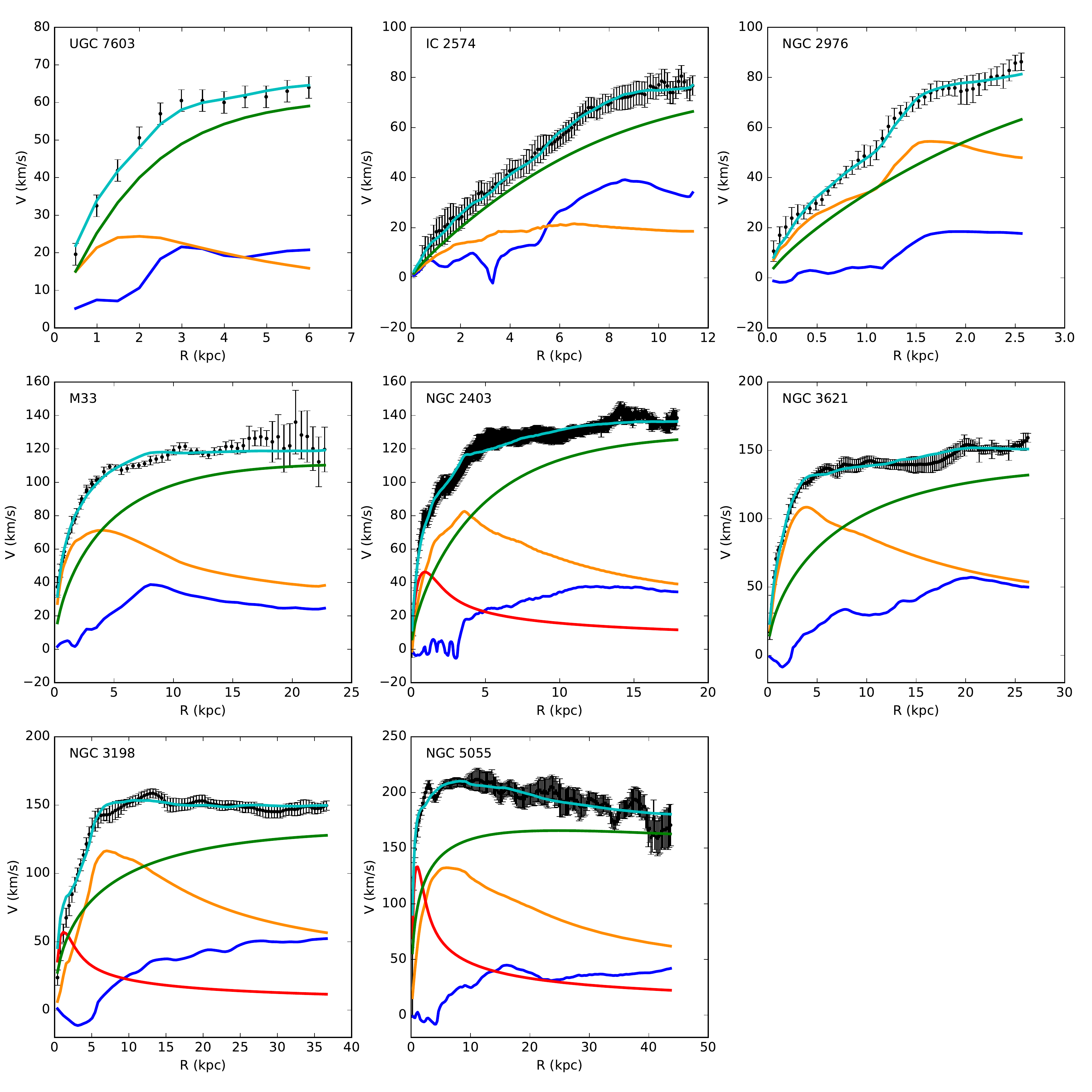}
\caption{continued.}
\end{figure*}
\par
\begin{figure*}[]
\includegraphics[width=\textwidth]{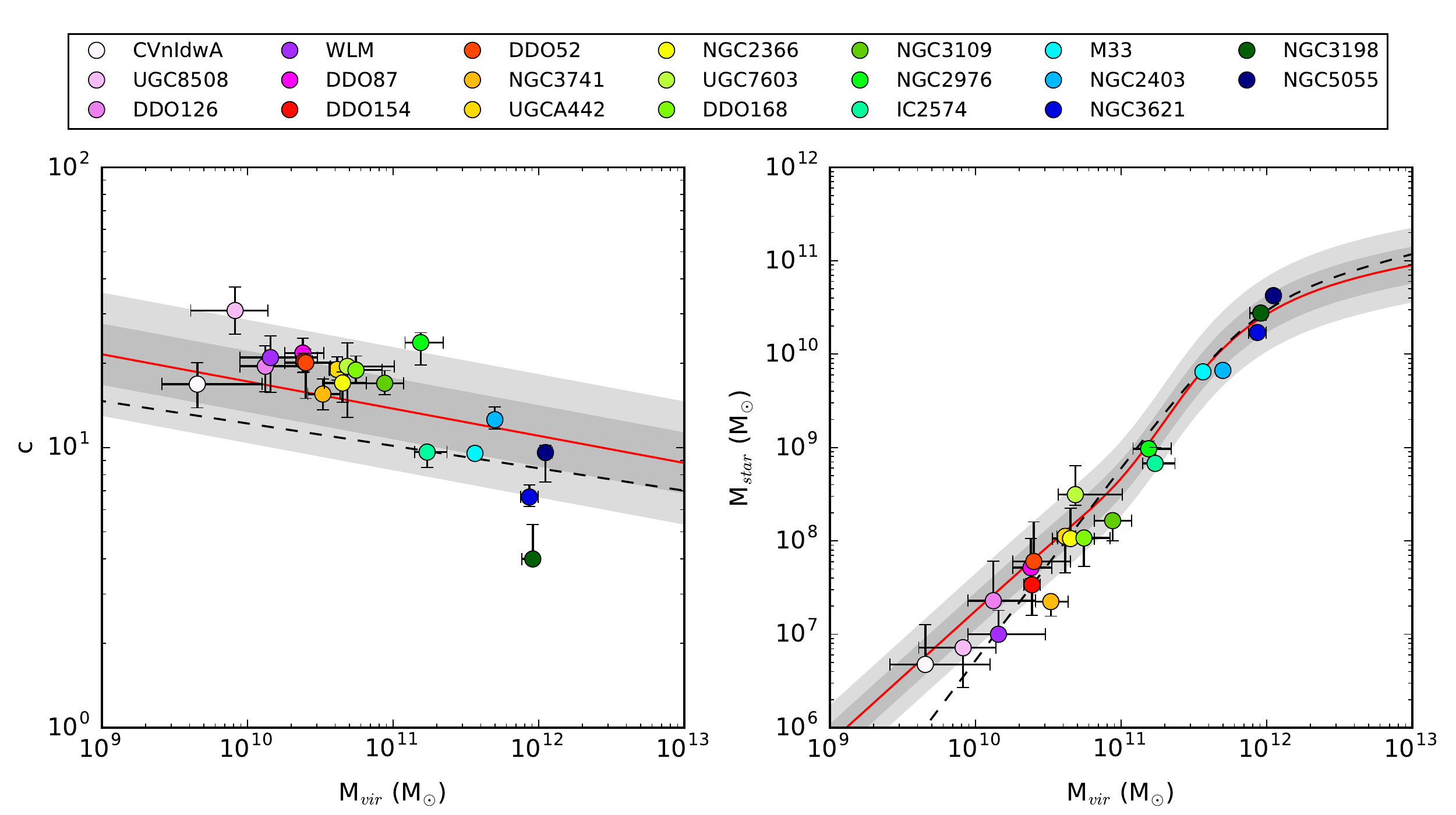}
\caption{Comparison of the parameters of the best-fit DC14 models to the cosmological halo mass-concentration relation from \cite{dutton14} (left) and the stellar mass-halo mass relation from \cite{behroozi13} (right). The error bars correspond to the extremal values of the multidimensional 68\% confidence region for each fit. The theoretical relations are shown as red lines and their 1$\sigma$ and 2$\sigma$ scatter are represented by the dark and light grey bands, respectively. The mass-concentration relation from \cite{maccio08} and the stellar mass-halo mass relation from \cite{behroozi13} are also shown as the black dashed lines.}
\label{fig:dc14_params}
\includegraphics[width=\textwidth]{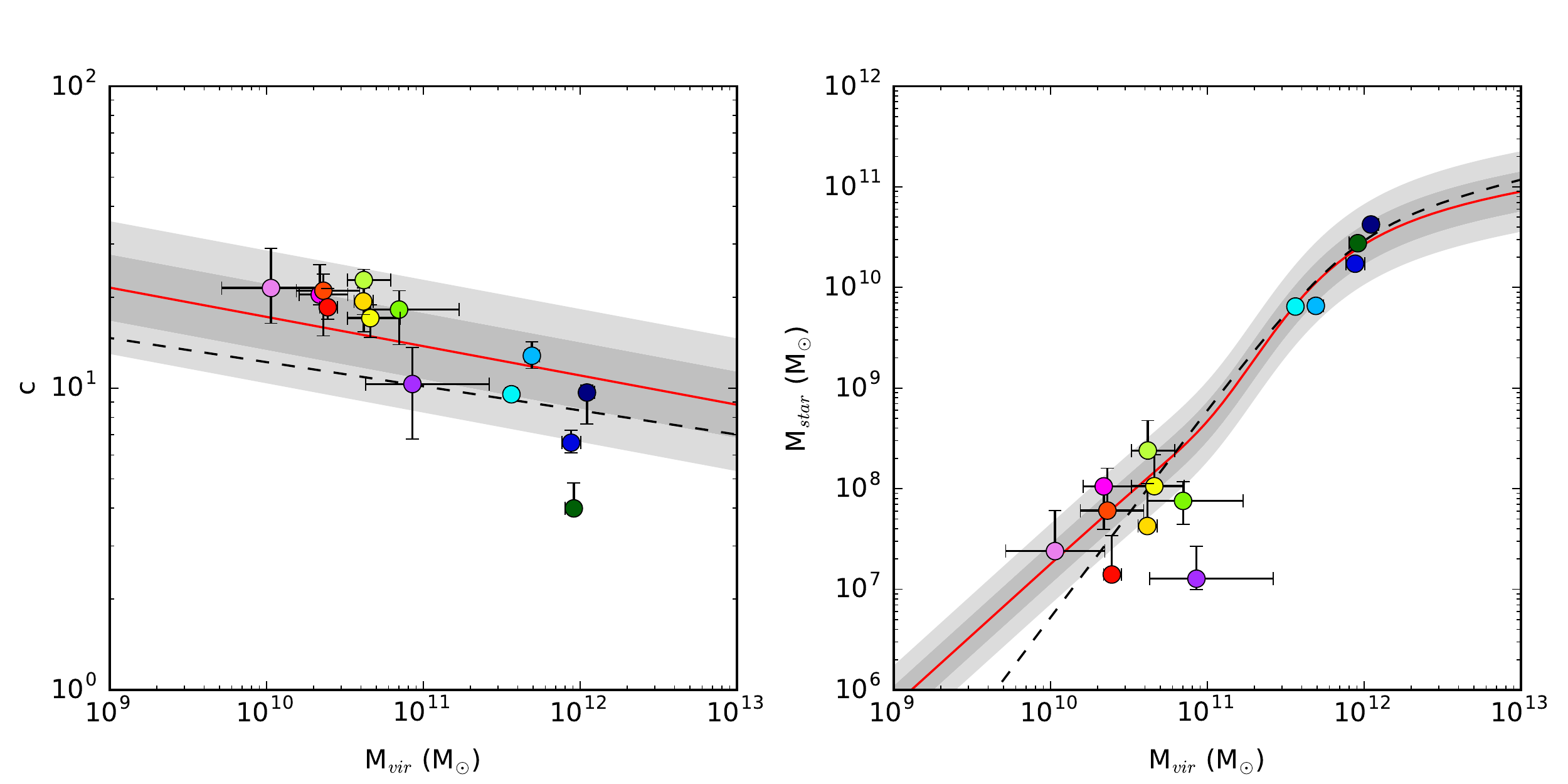}
\caption{Comparison of the parameters of the DC14 fits performed according to the strategy from \cite{pace16} to the cosmological halo mass-concentration relation from \cite{dutton14} (left) and the stellar mass-halo mass relation from \cite{behroozi13} (right). Colours are as in Figure \ref{fig:dc14_params}.}
\label{fig:dc14_limY}
\end{figure*}

In Figure \ref{fig:dc14_params} we compare the best-fit parameter values from our models with the cosmological halo mass-concentration and stellar mass-halo mass relations. These relations were derived from dark matter-only simulations (in combination with abundance matching). To account for this in the comparison, we scale our inferred halo masses as $M_{\text{vir}}$/(1-$f_{\text{b}}$), where $f_{\text{b}}$ is the Universal baryon fraction \citep[0.176 according to WMAP3;][]{mccarthy07}. The fits show excellent agreement with both scaling relations, although, somewhat surprisingly, our models seem to favour the $M_{\text{halo}}$-$c$ relation from \cite{dutton14} that is based on the Planck cosmology over the \cite{maccio08} relation that was used as prior in the fits. The only galaxy that falls significantly outside the 2$\sigma$ scatter of the \cite{dutton14} relation is NGC\,3198, but models with the concentration forced inside this scatter actually still provide a good fit to the data.
\par
Although the sample size is limited, our analysis thus seems to confirm the recent conclusion by \cite{katz16} that the DC14 halo is in good agreement with $\Lambda$CDM and can recover the predicted mass-concentration and stellar mass-halo mass relations in a sample of observed rotation curves. The latter might seem obvious since we use priors that `push' our fits towards these relations. The key point here, however, is that DC14 can produce dark matter haloes that \textit{simultaneously} provide good fits to the rotation curves and agree with the scaling relations; this is something that, for example the NFW halo, cannot do. The priors do not `make' physical solutions according to the scaling relations, but merely act as a filter to retain only the most physical solutions if they exist. The fact that our models prefer the \cite{dutton14} relation over the \cite{maccio08} relation that was used as prior is a good illustration of this.
\par
Our analysis contradicts the results from \cite{pace16}. We did not recover the huge range of halo concentrations that he found and we found no evidence for his claim that galaxies with $M_{*} \lesssim 10^{9}$ $M_{\odot}$ often reside in less massive haloes than predicted. It should be noted here that the modelling strategy from \cite{pace16} is somewhat different to our approach and that used by \cite{katz16}. \cite{pace16} used multi-nested sampling \citep[e.g.][]{feroz08} to fit his models and did not assume any physical priors between the parameters. The lack of priors results in posterior distributions that often contain multiple modes (or peaks). The mode with the lowest halo mass was selected as the final mode and galaxies for which the modes were too wide or not well separated were discarded. To investigate the effects of these differences, we performed a second set of fits without physical priors and used the criteria from \cite{pace16} to select the final models. The results are shown in Fig. \ref{fig:dc14_limY}. CVnIdwA, UGC\,8508, NGC\,3741, NGC\,3109, IC\,2571, and NGC\,2976 were discarded because their posterior distributions showed multiple blended modes; this was not the case in the original fits because the priors suppressed the additional modes. The best-fit model for WLM has shifted significantly. It is still consistent with the mass-concentration relation but now lies considerably below the stellar mass-halo mass relation. For the remaining galaxies the new models are essentially similar to the previous models or consistent within the uncertainties. The agreement with the scaling relations is still remarkably good, and although the number of galaxies is limited we find no evidence for the strong deviations that were reported by \cite{pace16}.

\subsection{CoreNFW}
\label{sec:cnfw_results}

\begin{figure*}[h!]
\includegraphics[width=\textwidth]{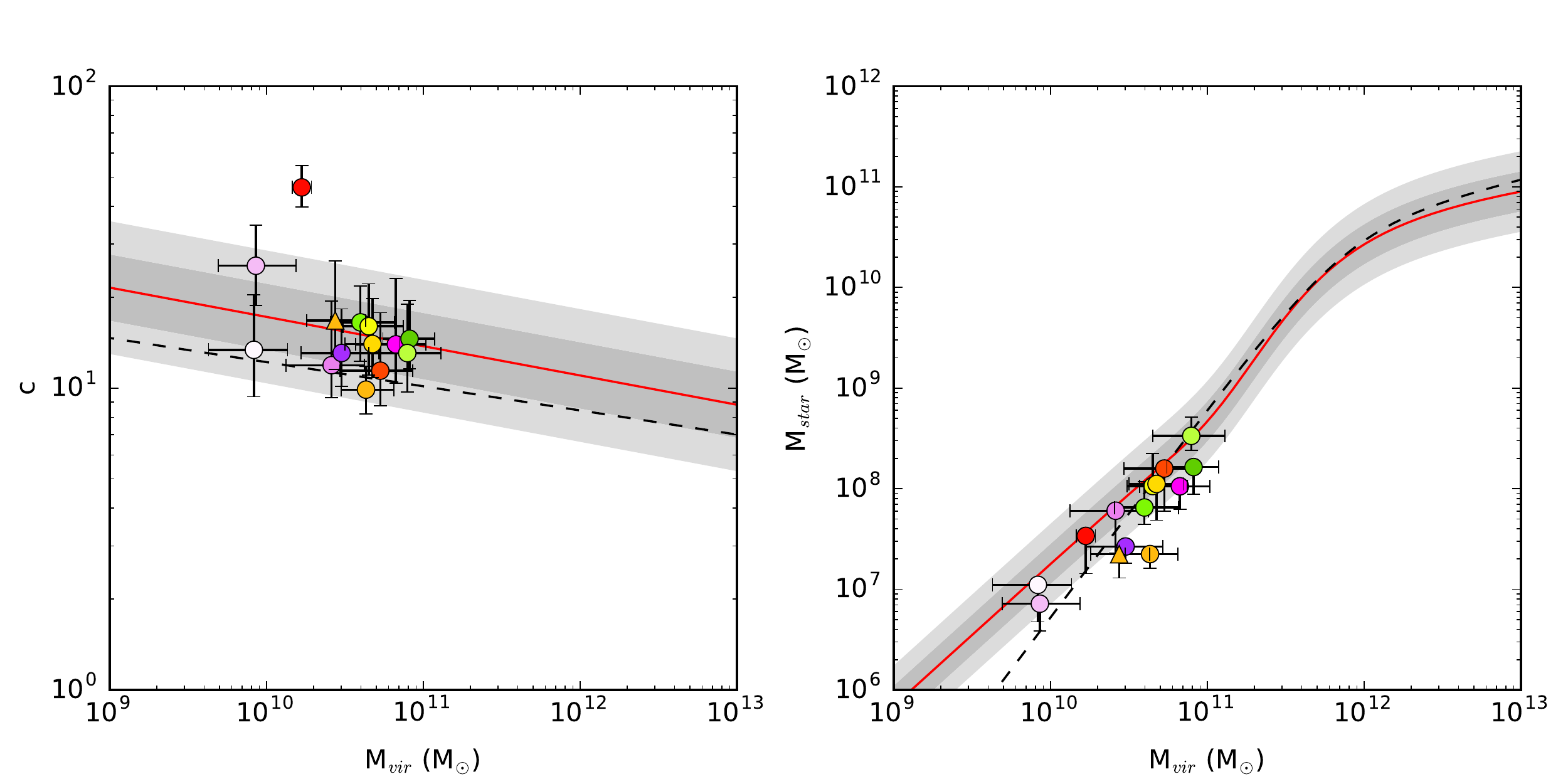}
\caption{Comparison of the parameters of the best-fit \textsc{core}NFW models to the cosmological halo mass-concentration relation from \cite{dutton14} (left) and the stellar mass-halo mass relation from \cite{behroozi13} (right). Colours are as in Figure \ref{fig:dc14_params}. The triangle represents the best-fit model for NGC\,3741 from a fit with the core radius as a free parameter (see section \ref{sec:differences}). The black dashed lines show the mass-concentration relation from \cite{maccio08} and the stellar mass-halo mass relation from \cite{moster10}.}
\label{fig:cnfw_params}
\end{figure*}

For the \textsc{core}NFW halo we limited the sample to the lower mass galaxies with $M_{\text{halo}} \lesssim 7 \times 10^{10}$ $M_{\odot}$. The individual \textsc{core}NFW fits for these galaxies are shown in Figure \ref{fig:cnfw_fixRc}. The models again provide a decent description of the data with no clearly bad fits except for DDO\,168. The fit results are compared to the cosmological scaling relations in Fig. \ref{fig:cnfw_params}. The agreement is very good and our models again seem to prefer the mass-concentration relation from \cite{dutton14} over that from \cite{maccio08}.
\par
The \textsc{core}NFW halo was also fitted to the same rotation curves by \cite{read16c}. Their results for the individual galaxies sometimes differ significantly from what is found in this work: \cite{read16c} have generally found somewhat lower halo and stellar masses, higher concentrations, and a better fit quality (lower $\chi^{2}_{\text{red}}$). These differences are, however, not unexpected. Indeed, while we derived the stellar and gas contributions from the measured surface density profiles from \cite{oh15}, \cite{read16c} used smooth exponential profiles based on \cite{zhang12} and \cite{oh15}. In addition, \cite{read16c} used the $M_{200}$ formalism while we used the \emph{virial} mass $M_{vir}$, so the halo parameters given in their Table 2 should not be compared directly to our values in Table \ref{table:results}. For an identical dark matter halo, our virial mass and concentration should be somewhat higher than the $M_{200}$ and $c_{200}$ values from \cite{read16c}. The fact that we generally find lower concentrations probably comes from the fact that \cite{read16c} did not use a mass-concentration prior, but instead set the boundaries for the concentration range based on the $M_{\text{vir}}$-$c$ relation from \cite{maccio07} and the extremities of their $M_{200}$ range. The rotation curve of NGC\,2366 that was reported by \cite{iorio16} also seems somewhat different from the curve that was used by (or at least shown in Fig. A2 of) \cite{read16c}.
\par
Despite these individual differences, the main conclusions remain the same. Both works generally find acceptable fits to the data and a good agreement with the stellar mass-halo mass relation.
\begin{figure*}[t!]
\includegraphics[width=\textwidth]{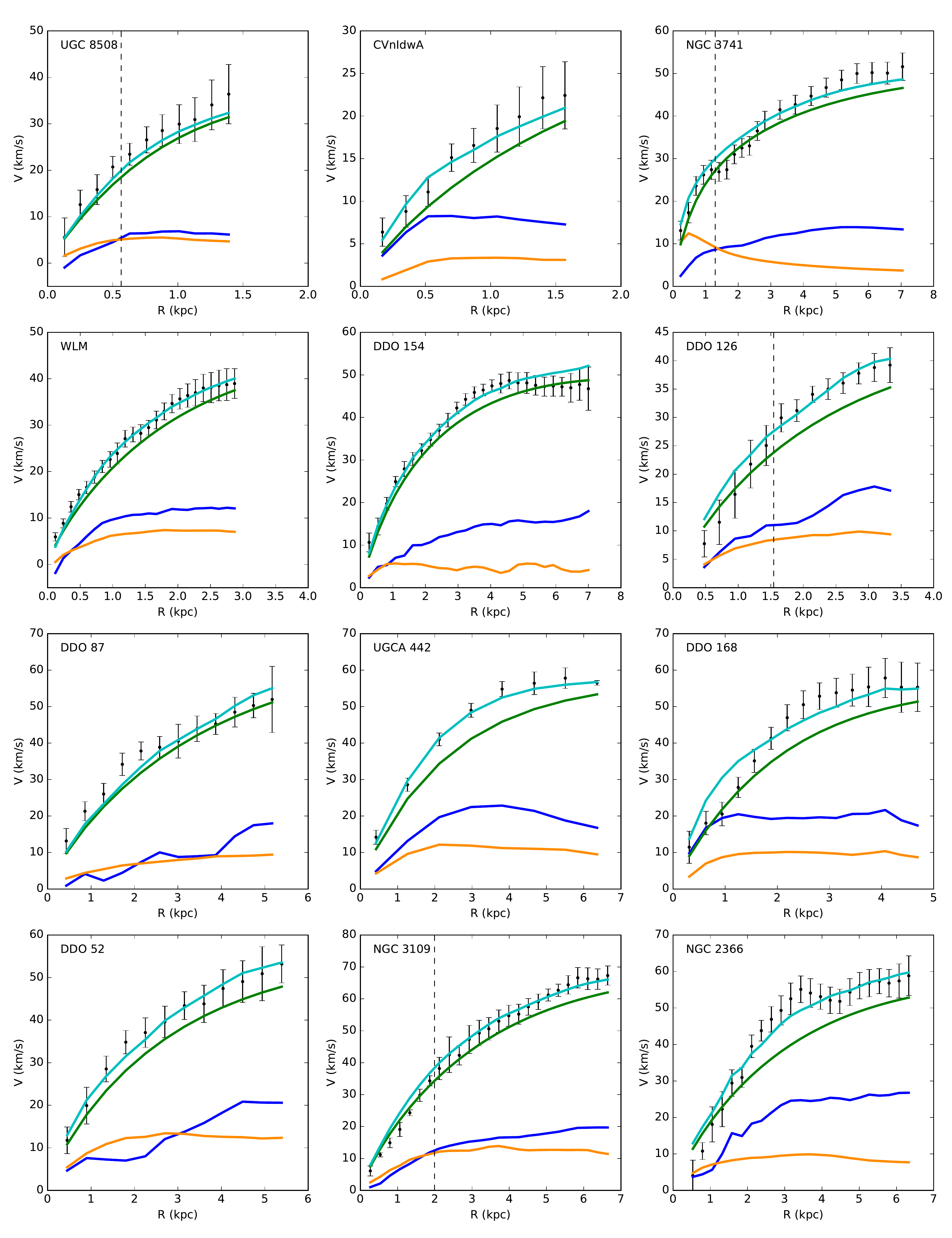}
\caption{Decomposition of the good rotation curves in our sample according to the maximum likelihood \textsc{core}NFW models. Colours and symbols are as in Figure \ref{fig:dc14_fits}.}
\label{fig:cnfw_fixRc}
\end{figure*}
\begin{figure}[]
\ContinuedFloat
\centering
\includegraphics[width=0.4\textwidth]{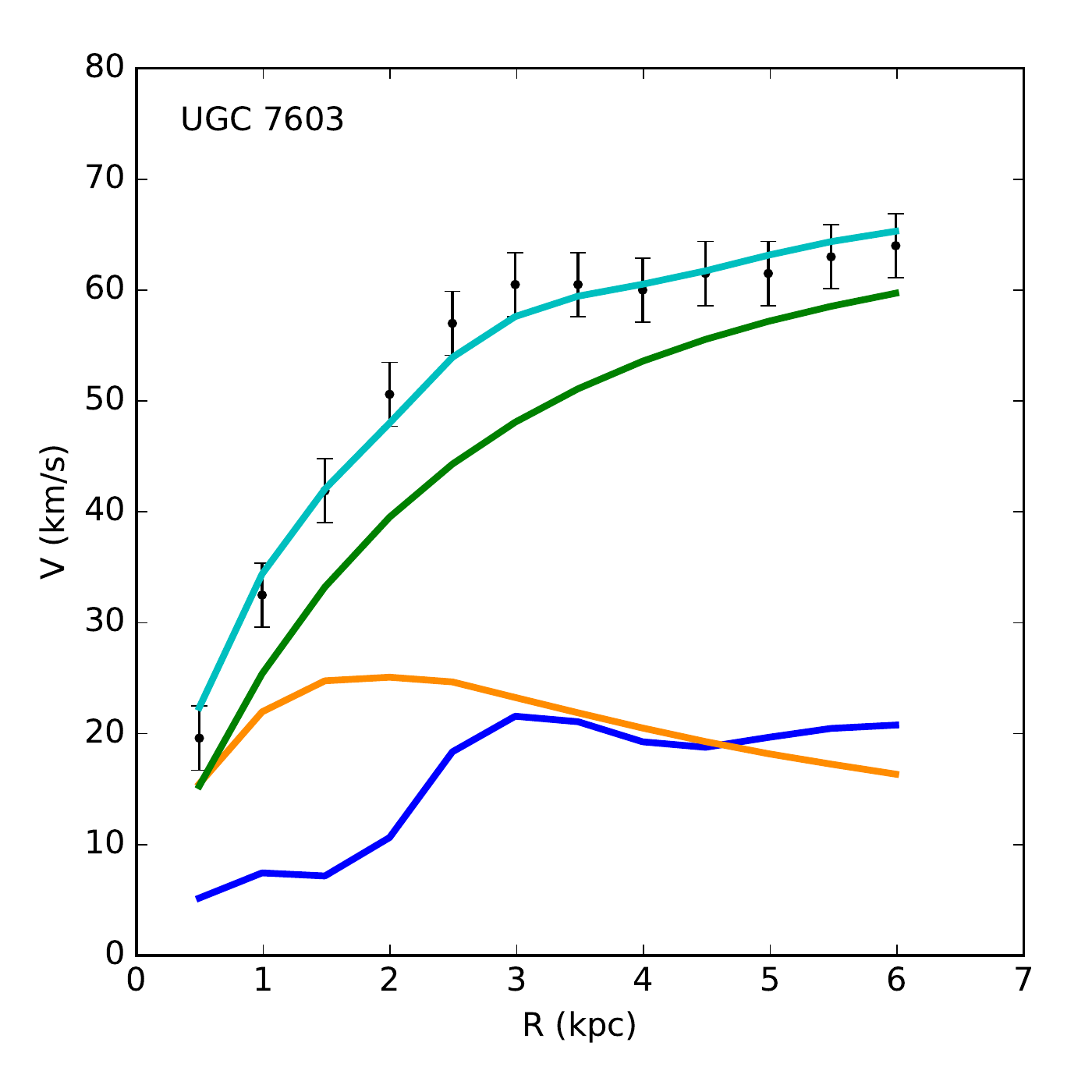}
\caption{continued.}
\end{figure}

\subsection{coreNFW versus DC14}
\label{sec:differences}

The \textsc{core}NFW halo model determines the strength of the dark matter core from the total time that the galaxy has been forming stars, while the radial extent of the core is related to the spatial distribution of the stars via the stellar half-mass radius. The stellar mass of the galaxy is not used. This is an important difference with the DC14 profile. On the one hand, DC14 uses $M_{*}$ instead of $t_{\text{SF}}$ as a measure of the amount of supernova feedback energy that has become available to form a core. On the other hand, this model also uses the additional gravitational potential due to $M_{*}$ as a mechanism to counteract core formation. In addition, the \textsc{core}NFW profile is essentially a pure NFW profile with its inner part flattened by feedback, while for DC14 the entire shape of the profile (i.e. $\gamma$, $\alpha,$ and $\beta$) changes as a function of the stellar-to-halo mass ratio.
\par
Given these rather different approaches, it is interesting to investigate how the dark matter haloes inferred by the \textsc{core}NFW and DC14 halo models compare across our sample. Since the physical mechanism that drives core formation is essentially the same for both models (supernova feedback after bursts of star formation), they should in principle give similar results in the overlapping halo mass range where they are both appropriate.

\subsubsection{Fit quality}

From Table \ref{table:results} we see that the DC14 and \textsc{core}NFW models generally give a similar fit quality. The DC14 model typically has a slightly lower $\chi^{2}_{\text{red}}$, with only WLM and UGCA\,442 breaking this trend, but in most cases both models represent the data well enough and it is not meaningful to classify one as better than the other. The latter also holds for DDO\,168, which is poorly fitted by both models. As discussed in Section \ref{sec:dc14_results} this is probably caused by a problem with the data.
\par
The only cases in which DC14 performs better than \textsc{core}NFW are NGC\,3741, DDO\,87, and DDO\,154. For NGC\,3741 and DDO\,87 the rotation curves are fitted significantly better by the DC14 model than by the \textsc{core}NFW model, although the latter is also still acceptable. For DDO\,154 the fit quality is good in both cases, but the \textsc{core}NFW halo needs an unphysically high concentration to achieve this. \textsc{core}NFW models with a lower concentration do not fit the data well. In each of these cases the problem seems to be related to the connection between the \textsc{core}NFW core radius and the stellar half-mass radius. For NGC\,3741 the rotation curve suggests an extended dark matter core, while the \textsc{core}NFW model has a cuspy NFW shape in all but the most central region because of the tiny half-mass radius of the stars. For the other two galaxies the rotation curves require a smaller core and a less `linear' dark matter contribution than derived from R$_{1/2}$.
\par
\begin{figure}[]
\centering
\includegraphics[width=0.4\textwidth]{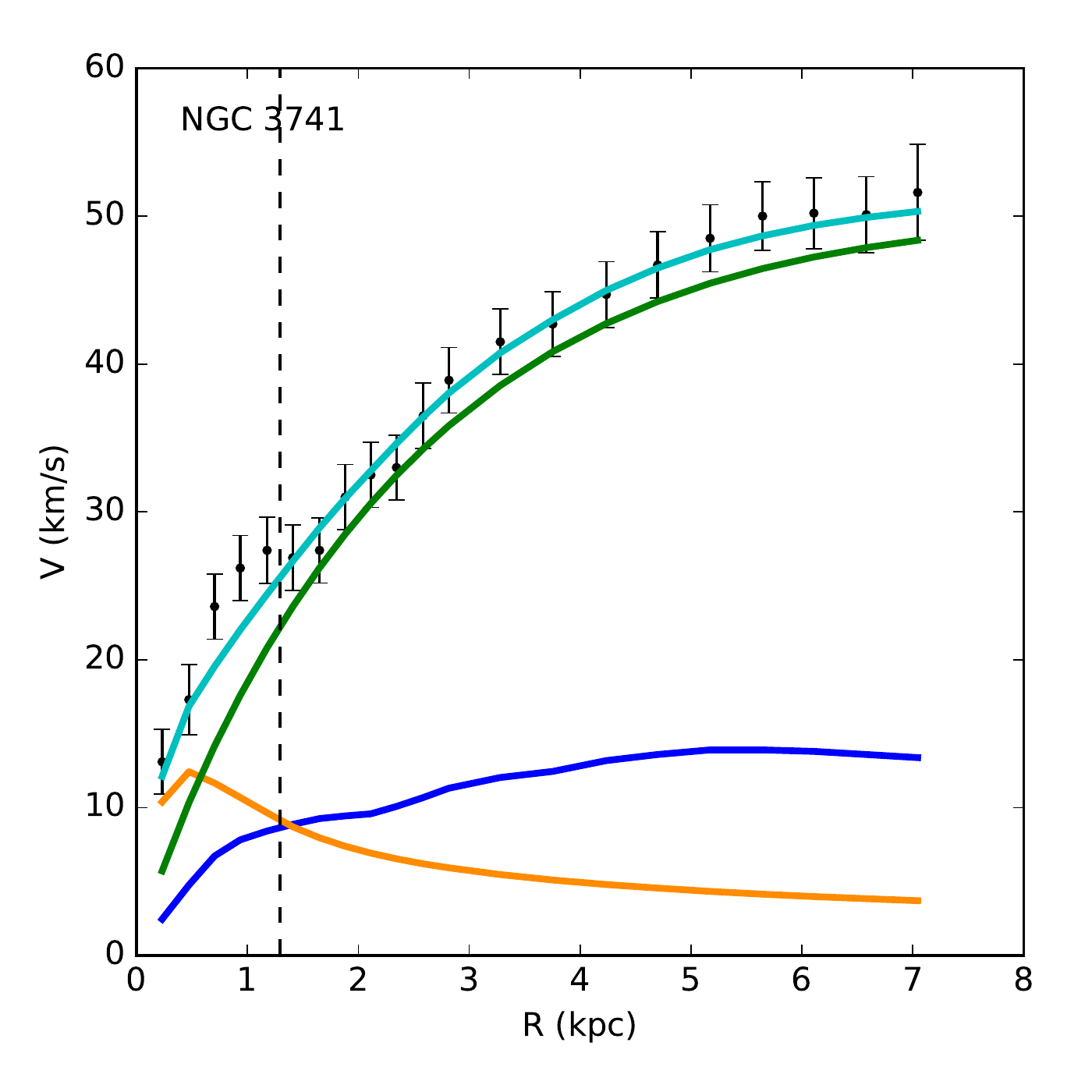}
\caption{Decomposition of the rotation curve of NGC\,3741 according to the best-fit \textsc{core}NFW model from a fit with the core radius as a free parameter. Colours and symbols are as in Figure \ref{fig:dc14_fits}.}
\label{fig:N3741_cnfw_free}
\end{figure}

These issues can be resolved by making $\eta$ a free parameter in the \textsc{core}NFW fits, but this would break with the prescription of \cite{read16a} and in a way make the comparison with the DC14 model unfair. For NGC\,3741 this might be justifiable. Indeed, while the stellar distribution of NGC\,3741 is unusually compact, the DC14 model finds sufficient stellar mass (i.e. supernova feedback energy) to form a substantial core that is in agreement with the data. Furthermore, the distribution of the gas is much more extended than that of the stars. Since the gravity of the outflowing gas after a supernova explosion drives the formation of a dark matter core, we could interpret the extended gas distribution in NGC\,3741 as a sign that the dark matter core radius is actually larger than that inferred from the half-mass radius of the stars. Therefore we performed a second \textsc{core}NFW fit for NGC\,3741 with $\eta$ as an additional free parameter (using a flat prior such that 0 < $r_{\text{c}}$ (kpc) < 7). The decomposed rotation curve is given in Figure \ref{fig:N3741_cnfw_free}. With a larger core radius of 4.06 kpc ($\eta = 11.2$) the \textsc{core}NFW model is now in excellent agreement with the data and with the DC14 model. This value of $\eta$ is significantly above the upper limit $\eta = 2.75$ derived by \cite{read16c}. However, this upper limit was derived under the assumption that $R_{\text{1/2}} \sim 0.015r_{200}$, which is also significantly larger than the value $R_{\text{1/2}} = 0.341$ kpc that is measured from the stellar distribution.
\par
For DDO\,154 and DDO\,87 the stellar distribution does not seem particularly unusual, although the inferred stellar half-mass radii are significantly larger by a factor of 3.9 and 1.8, respectively than those reported by \cite{read16c}. Nevertheless, this is also the case to some degree for CVnIdwA and DDO\,52 (factors of 1.6 and 1.4), for which DC14 and \textsc{core}NFW give similarly good fits. It is beyond the scope of this work to investigate whether this points to a problem with the \cite{oh15} data, and we limit ourselves to the conclusion that, based on the surface brightness profiles from \cite{oh15}, the DC14 model performs better than the \textsc{core}NFW model for the rotation curves of DDO\,154 and DDO\,87.
\begin{figure*}[]
\includegraphics[width=\textwidth]{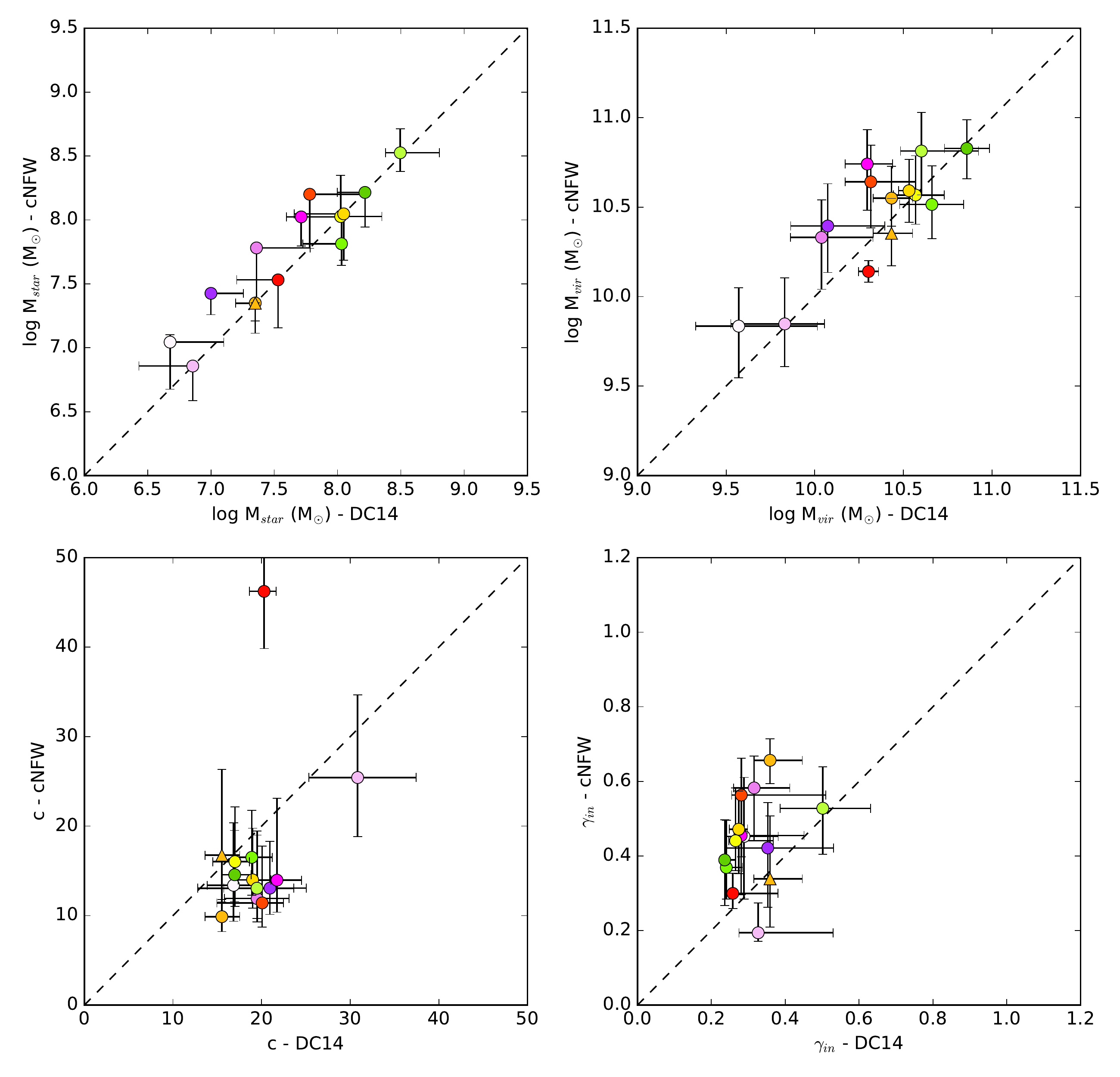}
\caption{Comparison of the main parameters of the best-fit DC14 and \textsc{core}NFW models: stellar mass (top left), halo mass (top right), halo concentration (bottom left), and log slope of the dark matter density at the innermost point of the rotation curve (bottom right). The inner log slopes in the latter plot were measured at the innermost point of each rotation curve and not in the actual centres of the galaxies. Therefore they are an upper limit to the real central slopes of the dark matter haloes. Colours and symbols are as in Figs \ref{fig:dc14_params} and \ref{fig:cnfw_params}.}
\label{fig:compare_pars}
\end{figure*}

\subsubsection{Best-fit parameters}

In Figure \ref{fig:compare_pars} we compare the main parameters of the best-fit DC14 and \textsc{core}NFW models for the galaxies in our sample with $M_{\text{halo}} \lesssim 7 \times 10^{10}$ $M_{\odot}$. With the exception of the unphysically high concentration of the \textsc{core}NFW model of DDO\,154, we find a fairly good agreement between the best-fit parameter values from both models. The stellar and virial masses follow the identity line, albeit with some scatter and the two models are generally consistent within the error bars. The DC14 model typically finds somewhat less extended cores (as can be seen by comparing the individual rotation curve decompositions) and more concentrated haloes, although except for DDO\,154 the concentrations are consistent within the errors. For the galaxies where the concentration difference is the highest, the DC14 model typically also finds somewhat lower stellar and halo masses than the \textsc{core}NFW model, which is consistent with the cosmological scaling relations. The bottom right panel in Fig. \ref{fig:compare_pars} shows the log slope of the dark matter density profiles measured at the innermost point of each rotation curve. This parameter strongly depends on the local shape of the density profile and should not be over-interpreted for the comparison between the two haloes. The main conclusion to be drawn is that both models find relatively cored dark matter haloes for all the galaxies with inner log slopes that are generally not too far apart and often consistent within the uncertainties. Unlike what is claimed by \cite{read16c}, we therefore find that the DC14 model can still show significant cusp-core transformations for galaxies with $M_{\text{halo}} \lesssim 10^{10}$ $M_{\odot}$.

\subsection{Problematic galaxies}

For completeness we show the individual fits to the problematic rotation curves in appendix \ref{app:problematic}. Using the flexible ($\alpha$,\,$\beta$,\,$\gamma$) profile, \cite{hague14} find good fits to the rotation curves of NGC\,925 and NGC\,7793 and also report tight constraints on the inner log slope of NGC\,3521 despite the poor fit of their model to the rising part of the rotation curve. In contrast, we find that the two physically motivated halo models investigated here cannot reproduce any of these rotation curves. For NGC\,3521 we even perform additional fits at the much smaller distance of 7.7 Mpc reported by SPARC, but with the same result. Finally the cuspy rotation curve of UGC\,8490 is well reproduced by the DC14 model, but \textsc{core}NFW strongly underestimates the rising part. However, for both these models the parameters lie well outside the cosmological scaling relations.

\section{Conclusions}
\label{sec:conclusions}

For a compact sample of 13 galaxies, spanning the mass range $M_{\text{halo}} \sim 4 \times 10^{9} - 7 \times 10^{10} M_{\odot}$, we have used MCMC to construct dynamical models of the rotation curves based on two recently proposed dark matter density profiles: the DC14 halo and \textsc{core}NFW halo. We further applied the DC14 halo to an additional set of higher mass galaxies with $M_{\text{halo}} \sim 10^{11} - 9 \times 10^{11} M_{\odot}$. The \textsc{core}NFW and DC14 halo models both use supernova feedback after bursts of star formation to transform primordial dark matter cusps into flatter cores. Although they were derived from simulations in complementary mass ranges ($M_{\text{halo}} \sim 10^{7} - 10^{9} M_{\odot}$ for \textsc{core}NFW versus $10^{10} - 8 \times 10^{11} M_{\odot}$ for DC14), both models should be valid for halo masses of $5 \times 10^{9}$ $M_{\odot} \lesssim M_{\text{halo}} \lesssim 5 \times 10^{10}$ $M_{\odot}$. With this analysis we investigated whether the \textsc{core}NFW and DC14 halo models converge to the same solutions in this overlapping mass range and whether their predictions agree with cosmological scaling relations.
\par
We found that both models are generally able to reproduce the rotation curves in our sample and find dark matter haloes that are in good agreement with the cosmological $M_{\text{halo}}$-$c$ and $M_{*}$-$M_{\text{halo}}$ relations, alleviating the cusp-core controversy. This confirms the results from \cite{read16c} for \textsc{core}NFW and from \cite{katz16} for DC14. On the other hand, we find no evidence of the huge scatter in concentrations or the disagreement of the DC14 predictions with the $M_{*}$-$M_{\text{halo}}$ relation that were recently claimed by \cite{pace16}, even if a similar modelling strategy is used.
\par
The two models generally give similarly good fits to the rotation curves, although the DC14 model does perform better in three cases. For NGC\,3741 and DDO\,87 the rotation curves are fitted significantly better by the DC14 model than by the \textsc{core}NFW model, although the latter is also still acceptable. For DDO\,154 the fit quality is good in both cases, but the \textsc{core}NFW halo needs an unphysically high concentration to achieve this. In each of these cases the problem for \textsc{core}NFW is related to the connection between the core size and the stellar half-mass radius.
\par
For NGC\,3741, a galaxy with a very compact stellar distribution but a remarkably extended atomic gas disk, we argue that it is justifiable to relax this connection and use $\eta$ as a free parameter in the \textsc{core}NFW fit. With a larger core radius the \textsc{core}NFW model is in excellent agreement with the data.
\par
The DC14 and \textsc{core}NFW haloes generally converge to (approximately) the same solution, as they should. Both models find cored dark matter haloes, and while DC14 tends to predict somewhat less extended cores and more concentrated haloes, the stellar masses, halo masses, and concentrations from both models are generally comparable and agree within the errors.

\begin{acknowledgements}

We thank the anonymous referee and the language editor for the useful comments and suggestions that significantly improved this paper. We also wish to thank E. de Blok for sharing his rotation curves and surface density profiles of the THINGS galaxies and D. Hunter and S.H. Oh for sharing the stellar and gas surface brightness profiles of the Little THINGS galaxies. F.A., M.B., and G.G. acknowledge the support of the Flemish Fund for Scientific Research (FWO-Vlaanderen). This research has made use of NASA's Astrophysics Data System and the NASA/IPAC Extragalactic Database (NED), which is operated by the Jet Propulsion Laboratory, California Institute of Technology, under contract with the National Aeronautics and Space Administration.

\end{acknowledgements}

\bibliography{Allaert_2017_biblio}{}

\begin{thebibliography}{79}
\expandafter\ifx\csname natexlab\endcsname\relax\def\natexlab#1{#1}\fi

\bibitem[{{Behroozi} {et~al.}(2013){Behroozi}, {Wechsler}, \&
  {Conroy}}]{behroozi13}
{Behroozi}, P.~S., {Wechsler}, R.~H., \& {Conroy}, C. 2013, \apj, 770, 57

\bibitem[{{Binney} \& {Tremaine}(2008)}]{binney08}
{Binney}, J. \& {Tremaine}, S. 2008, {Galactic Dynamics: Second Edition}
  (Princeton University Press)

\bibitem[{{Blais-Ouellette} {et~al.}(2001){Blais-Ouellette}, {Amram}, \&
  {Carignan}}]{blais01}
{Blais-Ouellette}, S., {Amram}, P., \& {Carignan}, C. 2001, \aj, 121, 1952

\bibitem[{{Broeils}(1992)}]{broeils92}
{Broeils}, A.~H. 1992, \aap, 256, 19

\bibitem[{{Corbelli} {et~al.}(2014){Corbelli}, {Thilker}, {Zibetti},
  {Giovanardi}, \& {Salucci}}]{corbelli14}
{Corbelli}, E., {Thilker}, D., {Zibetti}, S., {Giovanardi}, C., \& {Salucci},
  P. 2014, \aap, 572, A23

\bibitem[{{de Blok}(2010)}]{deblok10}
{de Blok}, W.~J.~G. 2010, Advances in Astronomy, 2010, 789293

\bibitem[{{de Blok} \& {Bosma}(2002)}]{deblok02}
{de Blok}, W.~J.~G. \& {Bosma}, A. 2002, \aap, 385, 816

\bibitem[{{de Blok} {et~al.}(2001){de Blok}, {McGaugh}, \& {Rubin}}]{deblok01}
{de Blok}, W.~J.~G., {McGaugh}, S.~S., \& {Rubin}, V.~C. 2001, \aj, 122, 2396

\bibitem[{{de Blok} {et~al.}(2008){de Blok}, {Walter}, {Brinks},
  {Trachternach}, {Oh}, \& {Kennicutt}}]{deblok08}
{de Blok}, W.~J.~G., {Walter}, F., {Brinks}, E., {et~al.} 2008, \aj, 136, 2648

\bibitem[{{Di Cintio} {et~al.}(2014{\natexlab{a}}){Di Cintio}, {Brook},
  {Dutton}, {Macci{\`o}}, {Stinson}, \& {Knebe}}]{dicintio14b}
{Di Cintio}, A., {Brook}, C.~B., {Dutton}, A.~A., {et~al.} 2014{\natexlab{a}},
  \mnras, 441, 2986

\bibitem[{{Di Cintio} {et~al.}(2014{\natexlab{b}}){Di Cintio}, {Brook},
  {Macci{\`o}}, {Stinson}, {Knebe}, {Dutton}, \& {Wadsley}}]{dicintio14a}
{Di Cintio}, A., {Brook}, C.~B., {Macci{\`o}}, A.~V., {et~al.}
  2014{\natexlab{b}}, \mnras, 437, 415

\bibitem[{{Diemand} {et~al.}(2005){Diemand}, {Zemp}, {Moore}, {Stadel}, \&
  {Carollo}}]{diemand05}
{Diemand}, J., {Zemp}, M., {Moore}, B., {Stadel}, J., \& {Carollo}, C.~M. 2005,
  \mnras, 364, 665

\bibitem[{{Dutton} \& {Macci{\`o}}(2014)}]{dutton14}
{Dutton}, A.~A. \& {Macci{\`o}}, A.~V. 2014, \mnras, 441, 3359

\bibitem[{{El-Zant} {et~al.}(2001){El-Zant}, {Shlosman}, \&
  {Hoffman}}]{elzant01}
{El-Zant}, A., {Shlosman}, I., \& {Hoffman}, Y. 2001, \apj, 560, 636

\bibitem[{{Feroz} \& {Hobson}(2008)}]{feroz08}
{Feroz}, F. \& {Hobson}, M.~P. 2008, \mnras, 384, 449

\bibitem[{{Foreman-Mackey} {et~al.}(2013){Foreman-Mackey}, {Hogg}, {Lang}, \&
  {Goodman}}]{foreman-mackey13}
{Foreman-Mackey}, D., {Hogg}, D.~W., {Lang}, D., \& {Goodman}, J. 2013, \pasp,
  125, 306

\bibitem[{{Freeman}(1970)}]{freeman70}
{Freeman}, K.~C. 1970, \apj, 160, 811

\bibitem[{{Gentile} {et~al.}(2005){Gentile}, {Burkert}, {Salucci}, {Klein}, \&
  {Walter}}]{gentile05}
{Gentile}, G., {Burkert}, A., {Salucci}, P., {Klein}, U., \& {Walter}, F. 2005,
  \apjl, 634, L145

\bibitem[{{Gentile} {et~al.}(2007){Gentile}, {Salucci}, {Klein}, \&
  {Granato}}]{gentile07}
{Gentile}, G., {Salucci}, P., {Klein}, U., \& {Granato}, G.~L. 2007, \mnras,
  375, 199

\bibitem[{{Gnedin} {et~al.}(2004){Gnedin}, {Kravtsov}, {Klypin}, \&
  {Nagai}}]{gnedin04}
{Gnedin}, O.~Y., {Kravtsov}, A.~V., {Klypin}, A.~A., \& {Nagai}, D. 2004, \apj,
  616, 16

\bibitem[{{Goodman} \& {Weare}(2010)}]{goodman10}
{Goodman}, J. \& {Weare}, J. 2010, Commun. Appl. Math. Comput. Sci., 5, 65

\bibitem[{{Governato} {et~al.}(2010){Governato}, {Brook}, {Mayer}, {Brooks},
  {Rhee}, {Wadsley}, {Jonsson}, {Willman}, {Stinson}, {Quinn}, \&
  {Madau}}]{governato10}
{Governato}, F., {Brook}, C., {Mayer}, L., {et~al.} 2010, \nat, 463, 203

\bibitem[{{Governato} {et~al.}(2012){Governato}, {Zolotov}, {Pontzen},
  {Christensen}, {Oh}, {Brooks}, {Quinn}, {Shen}, \& {Wadsley}}]{governato12}
{Governato}, F., {Zolotov}, A., {Pontzen}, A., {et~al.} 2012, \mnras, 422, 1231

\bibitem[{{Hague} \& {Wilkinson}(2013)}]{hague13}
{Hague}, P.~R. \& {Wilkinson}, M.~I. 2013, \mnras, 433, 2314

\bibitem[{{Hague} \& {Wilkinson}(2014)}]{hague14}
{Hague}, P.~R. \& {Wilkinson}, M.~I. 2014, \mnras, 443, 3712

\bibitem[{{Hague} \& {Wilkinson}(2015)}]{hague15}
{Hague}, P.~R. \& {Wilkinson}, M.~I. 2015, \apj, 800, 15

\bibitem[{Hastings(1970)}]{hastings70}
Hastings, W.~K. 1970, Biometrika, 57, 97

\bibitem[{{Hernquist}(1990)}]{hernquist90}
{Hernquist}, L. 1990, \apj, 356, 359

\bibitem[{{Hunter} {et~al.}(2012){Hunter}, {Ficut-Vicas}, {Ashley}, {Brinks},
  {Cigan}, {Elmegreen}, {Heesen}, {Herrmann}, {Johnson}, {Oh}, {Rupen},
  {Schruba}, {Simpson}, {Walter}, {Westpfahl}, {Young}, \& {Zhang}}]{hunter12}
{Hunter}, D.~A., {Ficut-Vicas}, D., {Ashley}, T., {et~al.} 2012, \aj, 144, 134

\bibitem[{{Iorio} {et~al.}(2017){Iorio}, {Fraternali}, {Nipoti}, {Di Teodoro},
  {Read}, \& {Battaglia}}]{iorio16}
{Iorio}, G., {Fraternali}, F., {Nipoti}, C., {et~al.} 2017, \mnras, 466, 4159

\bibitem[{{Jaffe}(1983)}]{jaffe83}
{Jaffe}, W. 1983, \mnras, 202, 995

\bibitem[{{Jesseit} {et~al.}(2002){Jesseit}, {Naab}, \& {Burkert}}]{jesseit02}
{Jesseit}, R., {Naab}, T., \& {Burkert}, A. 2002, \apjl, 571, L89

\bibitem[{{Katz} {et~al.}(2016){Katz}, {Lelli}, {McGaugh}, {Di Cintio},
  {Brook}, \& {Schombert}}]{katz16}
{Katz}, H., {Lelli}, F., {McGaugh}, S.~S., {et~al.} 2016, ArXiv e-prints

\bibitem[{{Kirichenko} {et~al.}(2015){Kirichenko}, {Danilenko}, {Shternin},
  {Shibanov}, {Ryspaeva}, {Zyuzin}, {Durant}, {Kargaltsev}, {Pavlov}, \&
  {Cabrera-Lavers}}]{kirichenko15}
{Kirichenko}, A., {Danilenko}, A., {Shternin}, P., {et~al.} 2015, \apj, 802, 17

\bibitem[{{Klypin} {et~al.}(2001){Klypin}, {Kravtsov}, {Bullock}, \&
  {Primack}}]{klypin01}
{Klypin}, A., {Kravtsov}, A.~V., {Bullock}, J.~S., \& {Primack}, J.~R. 2001,
  \apj, 554, 903

\bibitem[{{Klypin} {et~al.}(2011){Klypin}, {Trujillo-Gomez}, \&
  {Primack}}]{klypin11}
{Klypin}, A.~A., {Trujillo-Gomez}, S., \& {Primack}, J. 2011, \apj, 740, 102

\bibitem[{{Lelli} {et~al.}(2016){Lelli}, {McGaugh}, \& {Schombert}}]{lelli16}
{Lelli}, F., {McGaugh}, S.~S., \& {Schombert}, J.~M. 2016, ArXiv e-prints

\bibitem[{{Macci{\`o}} {et~al.}(2008){Macci{\`o}}, {Dutton}, \& {van den
  Bosch}}]{maccio08}
{Macci{\`o}}, A.~V., {Dutton}, A.~A., \& {van den Bosch}, F.~C. 2008, \mnras,
  391, 1940

\bibitem[{{Macci{\`o}} {et~al.}(2007){Macci{\`o}}, {Dutton}, {van den Bosch},
  {Moore}, {Potter}, \& {Stadel}}]{maccio07}
{Macci{\`o}}, A.~V., {Dutton}, A.~A., {van den Bosch}, F.~C., {et~al.} 2007,
  \mnras, 378, 55

\bibitem[{{Macci{\`o}} {et~al.}(2012){Macci{\`o}}, {Stinson}, {Brook},
  {Wadsley}, {Couchman}, {Shen}, {Gibson}, \& {Quinn}}]{maccio12}
{Macci{\`o}}, A.~V., {Stinson}, G., {Brook}, C.~B., {et~al.} 2012, \apjl, 744,
  L9

\bibitem[{{McCarthy} {et~al.}(2007){McCarthy}, {Bower}, \&
  {Balogh}}]{mccarthy07}
{McCarthy}, I.~G., {Bower}, R.~G., \& {Balogh}, M.~L. 2007, \mnras, 377, 1457

\bibitem[{{McGaugh} \& {Schombert}(2014)}]{mcgaughschombert14}
{McGaugh}, S.~S. \& {Schombert}, J.~M. 2014, \aj, 148, 77

\bibitem[{{McQuinn} {et~al.}(2015){McQuinn}, {Lelli}, {Skillman}, {Dolphin},
  {McGaugh}, \& {Williams}}]{mcquinn15}
{McQuinn}, K.~B.~W., {Lelli}, F., {Skillman}, E.~D., {et~al.} 2015, ArXiv
  e-prints

\bibitem[{{Meidt} {et~al.}(2014){Meidt}, {Schinnerer}, {van de Ven},
  {Zaritsky}, {Peletier}, {Knapen}, {Sheth}, {Regan}, {Querejeta},
  {Mu{\~n}oz-Mateos}, {Kim}, {Hinz}, {Gil de Paz}, {Athanassoula}, {Bosma},
  {Buta}, {Cisternas}, {Ho}, {Holwerda}, {Skibba}, {Laurikainen}, {Salo},
  {Gadotti}, {Laine}, {Erroz-Ferrer}, {Comer{\'o}n}, {Men{\'e}ndez-Delmestre},
  {Seibert}, \& {Mizusawa}}]{meidt14}
{Meidt}, S.~E., {Schinnerer}, E., {van de Ven}, G., {et~al.} 2014, \apj, 788,
  144

\bibitem[{Metropolis {et~al.}(1953)Metropolis, Rosenbluth, Rosenbluth, Teller,
  \& Teller}]{metropolis53}
Metropolis, N., Rosenbluth, A., Rosenbluth, M., Teller, A., \& Teller, E. 1953,
  J. Chem. Phys., 21, 1087

\bibitem[{{Moore} {et~al.}(1999){Moore}, {Quinn}, {Governato}, {Stadel}, \&
  {Lake}}]{moore99}
{Moore}, B., {Quinn}, T., {Governato}, F., {Stadel}, J., \& {Lake}, G. 1999,
  \mnras, 310, 1147

\bibitem[{{Moster} {et~al.}(2010){Moster}, {Somerville}, {Maulbetsch}, {van den
  Bosch}, {Macci{\`o}}, {Naab}, \& {Oser}}]{moster10}
{Moster}, B.~P., {Somerville}, R.~S., {Maulbetsch}, C., {et~al.} 2010, \apj,
  710, 903

\bibitem[{{Navarro} {et~al.}(1996{\natexlab{a}}){Navarro}, {Eke}, \&
  {Frenk}}]{navarro96b}
{Navarro}, J.~F., {Eke}, V.~R., \& {Frenk}, C.~S. 1996{\natexlab{a}}, \mnras,
  283, L72

\bibitem[{{Navarro} {et~al.}(1996{\natexlab{b}}){Navarro}, {Frenk}, \&
  {White}}]{navarro96}
{Navarro}, J.~F., {Frenk}, C.~S., \& {White}, S.~D.~M. 1996{\natexlab{b}},
  \apj, 462, 563

\bibitem[{{O{\~n}orbe} {et~al.}(2015){O{\~n}orbe}, {Boylan-Kolchin}, {Bullock},
  {Hopkins}, {Kere{\v s}}, {Faucher-Gigu{\`e}re}, {Quataert}, \&
  {Murray}}]{onorbe15}
{O{\~n}orbe}, J., {Boylan-Kolchin}, M., {Bullock}, J.~S., {et~al.} 2015,
  \mnras, 454, 2092

\bibitem[{{Oh} {et~al.}(2011){Oh}, {de Blok}, {Brinks}, {Walter}, \&
  {Kennicutt}}]{oh11}
{Oh}, S.-H., {de Blok}, W.~J.~G., {Brinks}, E., {Walter}, F., \& {Kennicutt},
  Jr., R.~C. 2011, \aj, 141, 193

\bibitem[{{Oh} {et~al.}(2015){Oh}, {Hunter}, {Brinks}, {Elmegreen}, {Schruba},
  {Walter}, {Rupen}, {Young}, {Simpson}, {Johnson}, {Herrmann}, {Ficut-Vicas},
  {Cigan}, {Heesen}, {Ashley}, \& {Zhang}}]{oh15}
{Oh}, S.-H., {Hunter}, D.~A., {Brinks}, E., {et~al.} 2015, \aj, 149, 180

\bibitem[{{Oort}(1932)}]{oort32}
{Oort}, J.~H. 1932, \bain, 6, 249

\bibitem[{{Pace}(2016)}]{pace16}
{Pace}, A.~B. 2016, ArXiv e-prints

\bibitem[{{Pineda} {et~al.}(2016){Pineda}, {Hayward}, {Springel}, \& {Mendes de
  Oliveira}}]{pineda16}
{Pineda}, J.~C.~B., {Hayward}, C.~C., {Springel}, V., \& {Mendes de Oliveira},
  C. 2016, ArXiv e-prints

\bibitem[{Press {et~al.}(2007)Press, Teukolsky, Vetterling, \&
  Flannery}]{press07}
Press, W.~H., Teukolsky, S.~A., Vetterling, W.~T., \& Flannery, B.~P. 2007,
  Numerical Recipes 3rd Edition: The Art of Scientific Computing, 3rd edn. (New
  York, NY, USA: Cambridge University Press)

\bibitem[{{Puglielli} {et~al.}(2010){Puglielli}, {Widrow}, \&
  {Courteau}}]{puglielli10}
{Puglielli}, D., {Widrow}, L.~M., \& {Courteau}, S. 2010, \apj, 715, 1152

\bibitem[{{Read} {et~al.}(2016{\natexlab{a}}){Read}, {Agertz}, \&
  {Collins}}]{read16a}
{Read}, J.~I., {Agertz}, O., \& {Collins}, M.~L.~M. 2016{\natexlab{a}}, \mnras,
  459, 2573

\bibitem[{{Read} \& {Gilmore}(2005)}]{read05}
{Read}, J.~I. \& {Gilmore}, G. 2005, \mnras, 356, 107

\bibitem[{{Read} {et~al.}(2016{\natexlab{b}}){Read}, {Iorio}, {Agertz}, \&
  {Fraternali}}]{read16c}
{Read}, J.~I., {Iorio}, G., {Agertz}, O., \& {Fraternali}, F.
  2016{\natexlab{b}}, ArXiv e-prints

\bibitem[{{Read} {et~al.}(2016{\natexlab{c}}){Read}, {Iorio}, {Agertz}, \&
  {Fraternali}}]{read16b}
{Read}, J.~I., {Iorio}, G., {Agertz}, O., \& {Fraternali}, F.
  2016{\natexlab{c}}, \mnras, 462, 3628

\bibitem[{{Rhee} {et~al.}(2004){Rhee}, {Valenzuela}, {Klypin}, {Holtzman}, \&
  {Moorthy}}]{rhee04}
{Rhee}, G., {Valenzuela}, O., {Klypin}, A., {Holtzman}, J., \& {Moorthy}, B.
  2004, \apj, 617, 1059

\bibitem[{{Roberts}(1975)}]{roberts75}
{Roberts}, M.~S. 1975, in IAU Symposium, Vol.~69, Dynamics of the Solar
  Systems, ed. A.~{Hayli}, 331

\bibitem[{{Romano-D{\'{\i}}az} {et~al.}(2008){Romano-D{\'{\i}}az}, {Shlosman},
  {Hoffman}, \& {Heller}}]{romano08}
{Romano-D{\'{\i}}az}, E., {Shlosman}, I., {Hoffman}, Y., \& {Heller}, C. 2008,
  \apjl, 685, L105

\bibitem[{{Rubin} {et~al.}(1978){Rubin}, {Thonnard}, \& {Ford}}]{rubin78}
{Rubin}, V.~C., {Thonnard}, N., \& {Ford}, Jr., W.~K. 1978, \apjl, 225, L107

\bibitem[{{Spekkens} {et~al.}(2005){Spekkens}, {Giovanelli}, \&
  {Haynes}}]{spekkens05}
{Spekkens}, K., {Giovanelli}, R., \& {Haynes}, M.~P. 2005, \aj, 129, 2119

\bibitem[{{Spergel} {et~al.}(2007){Spergel}, {Bean}, {Dor{\'e}}, {Nolta},
  {Bennett}, {Dunkley}, {Hinshaw}, {Jarosik}, {Komatsu}, {Page}, {Peiris},
  {Verde}, {Halpern}, {Hill}, {Kogut}, {Limon}, {Meyer}, {Odegard}, {Tucker},
  {Weiland}, {Wollack}, \& {Wright}}]{spergel07}
{Spergel}, D.~N., {Bean}, R., {Dor{\'e}}, O., {et~al.} 2007, \apjs, 170, 377

\bibitem[{{Stadel} {et~al.}(2009){Stadel}, {Potter}, {Moore}, {Diemand},
  {Madau}, {Zemp}, {Kuhlen}, \& {Quilis}}]{stadel09}
{Stadel}, J., {Potter}, D., {Moore}, B., {et~al.} 2009, \mnras, 398, L21

\bibitem[{{Swaters} {et~al.}(2003){Swaters}, {Madore}, {van den Bosch}, \&
  {Balcells}}]{swaters03}
{Swaters}, R.~A., {Madore}, B.~F., {van den Bosch}, F.~C., \& {Balcells}, M.
  2003, \apj, 583, 732

\bibitem[{{Teyssier} {et~al.}(2013){Teyssier}, {Pontzen}, {Dubois}, \&
  {Read}}]{teyssier13}
{Teyssier}, R., {Pontzen}, A., {Dubois}, Y., \& {Read}, J.~I. 2013, \mnras,
  429, 3068

\bibitem[{{Tonini} {et~al.}(2006){Tonini}, {Lapi}, \& {Salucci}}]{tonini06}
{Tonini}, C., {Lapi}, A., \& {Salucci}, P. 2006, \apj, 649, 591

\bibitem[{{Tully} {et~al.}(2013){Tully}, {Courtois}, {Dolphin}, {Fisher},
  {H{\'e}raudeau}, {Jacobs}, {Karachentsev}, {Makarov}, {Makarova},
  {Mitronova}, {Rizzi}, {Shaya}, {Sorce}, \& {Wu}}]{tully13}
{Tully}, R.~B., {Courtois}, H.~M., {Dolphin}, A.~E., {et~al.} 2013, \aj, 146,
  86

\bibitem[{{Valenzuela} {et~al.}(2007){Valenzuela}, {Rhee}, {Klypin},
  {Governato}, {Stinson}, {Quinn}, \& {Wadsley}}]{valenzuela07}
{Valenzuela}, O., {Rhee}, G., {Klypin}, A., {et~al.} 2007, \apj, 657, 773

\bibitem[{{van Albada} {et~al.}(1985){van Albada}, {Bahcall}, {Begeman}, \&
  {Sancisi}}]{vanalbada85}
{van Albada}, T.~S., {Bahcall}, J.~N., {Begeman}, K., \& {Sancisi}, R. 1985,
  \apj, 295, 305

\bibitem[{{Walter} {et~al.}(2008){Walter}, {Brinks}, {de Blok}, {Bigiel},
  {Kennicutt}, {Thornley}, \& {Leroy}}]{walter08}
{Walter}, F., {Brinks}, E., {de Blok}, W.~J.~G., {et~al.} 2008, \aj, 136, 2563

\bibitem[{{Weldrake} {et~al.}(2003){Weldrake}, {de Blok}, \&
  {Walter}}]{weldrake03}
{Weldrake}, D.~T.~F., {de Blok}, W.~J.~G., \& {Walter}, F. 2003, \mnras, 340,
  12

\bibitem[{{Zhang} {et~al.}(2012){Zhang}, {Hunter}, {Elmegreen}, {Gao}, \&
  {Schruba}}]{zhang12}
{Zhang}, H.-X., {Hunter}, D.~A., {Elmegreen}, B.~G., {Gao}, Y., \& {Schruba},
  A. 2012, \aj, 143, 47

\bibitem[{{Zhao}(1996)}]{zhao96}
{Zhao}, H. 1996, \mnras, 278, 488

\bibitem[{{Zwicky}(1933)}]{zwicky33}
{Zwicky}, F. 1933, Helvetica Physica Acta, 6, 110

\end{thebibliography}
\bibliographystyle{aa}

\begin{appendix}

\section{Problematic rotation curves}
\label{app:problematic}

\begin{figure*}[h!]
\includegraphics[width=\textwidth]{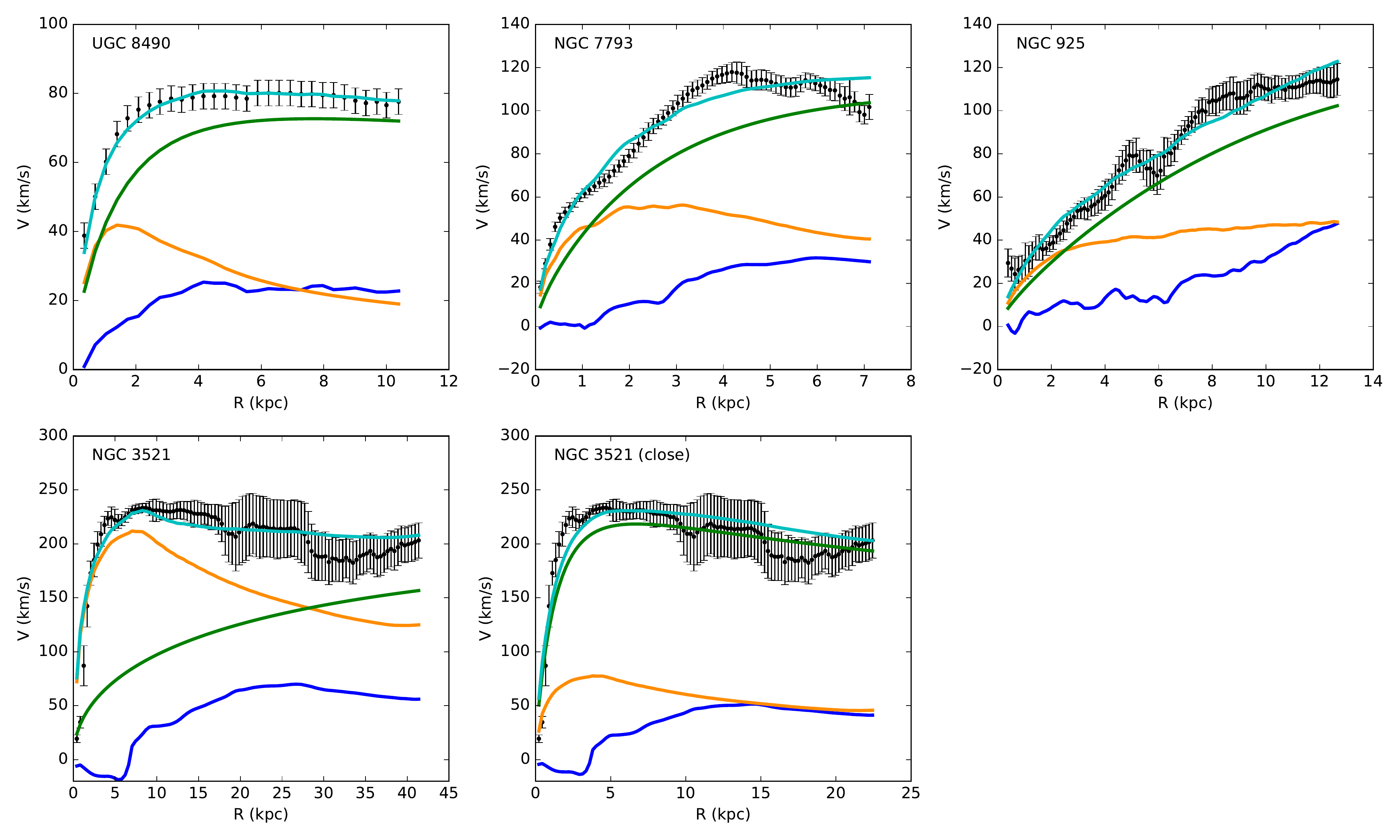}
\caption{Decomposition of the problematic rotation curves in our sample according to the best-fit DC14 models. Colors and symbols are as in Figure \ref{fig:dc14_fits}.}
\label{fig:dc14_rogue}
\end{figure*}

\begin{figure*}[b!]
\includegraphics[width=\textwidth]{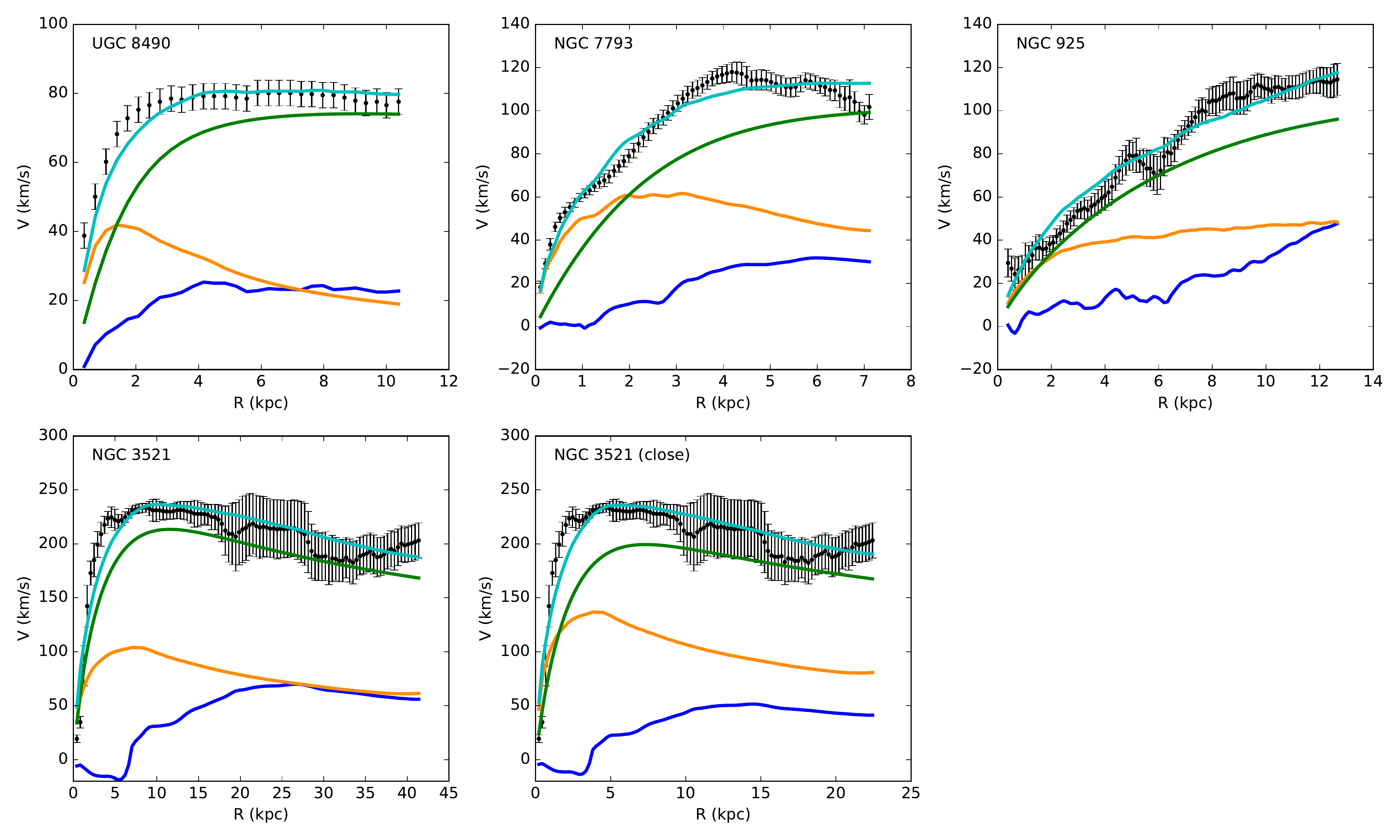}
\caption{Decomposition of the problematic rotation curves in our sample according to the best-fit \textsc{core}NFW models. Colors and symbols are as in Figure \ref{fig:dc14_fits}.}
\label{fig:cnfw_rogue}
\end{figure*}

\end{appendix}

\end{document}